\documentclass[%
prb,
twocolumn,
amsmath,
amssymb,
reprint,
superscriptaddress,
footinbib,
amsmath,amssymb,
aps,
]{revtex4}

\usepackage{graphicx}% Include figure files
\usepackage{dcolumn}% Align table columns on decimal point
\usepackage{bm}% bold math
\usepackage{natbib}
\usepackage{xcolor}
\usepackage{soul}
\usepackage{upgreek}
\usepackage[]{changes}
\usepackage{color}
\usepackage{hyperref}
\hypersetup{
	colorlinks,
	allcolors=blue,
	linktoc=all
}
\definechangesauthor[name=MA, color=purple]{Miles}

\newcommand*\SiN{Si$_3$N$_4$ }

\begin{document}
\preprint{APS/123-QED}
\title{Zero-dispersion Kerr solitons in optical microresonators}

\author{Miles H. Anderson}
\affiliation{Institute of Physics, Swiss Federal Institute of Technology (EPFL), Lausanne, Switzerland}

\author{Grigory Lihachev}
\affiliation{Institute of Physics, Swiss Federal Institute of Technology (EPFL), Lausanne, Switzerland}

\author{Wenle Weng}
\affiliation{Institute of Physics, Swiss Federal Institute of Technology (EPFL), Lausanne, Switzerland}

\author{Junqiu Liu}
\affiliation{Institute of Physics, Swiss Federal Institute of Technology (EPFL), Lausanne, Switzerland}

\author{Tobias J. Kippenberg}
\email{tobias.kippenberg@epfl.ch}
\affiliation{Institute of Physics, Swiss Federal Institute of Technology (EPFL), Lausanne, Switzerland}

\pacs{Valid PACS appear here}% PACS, the Physics and Astronomy
\maketitle

\begin{figure*} [t]
	\includegraphics{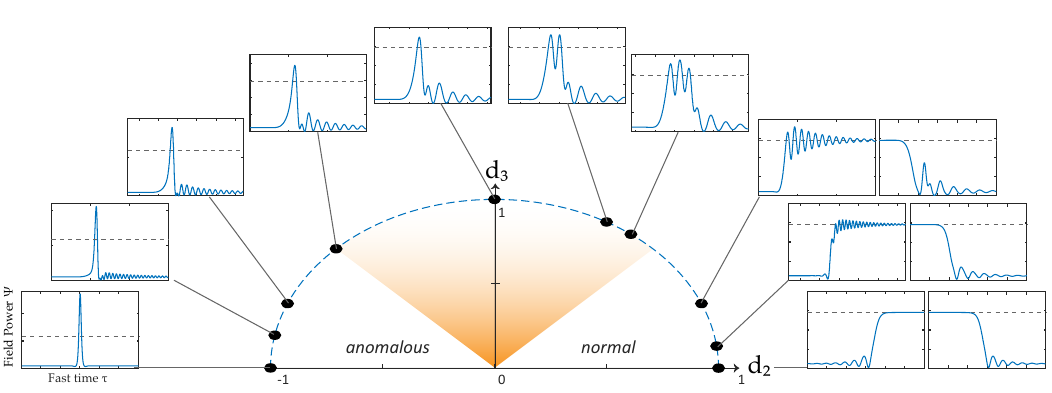}
	\caption{
		\textbf{Localized Dissipative Structures in the second-/third-order dispersion ($\mathbf{d_2/d_3}$) plane. } Clockwise from left: conventional dissipative solitons, dissipative solitons with dispersive-wave tails, zero-dispersion solitons (orange area), switching waves with dispersive wave tails, conventional switching wave. Dashed gray line in figures represent the CW high-state solution. Video available as supplementary info.}
	\label{fig:concept01}
\end{figure*}

\begin{figure*} [t]
	\includegraphics{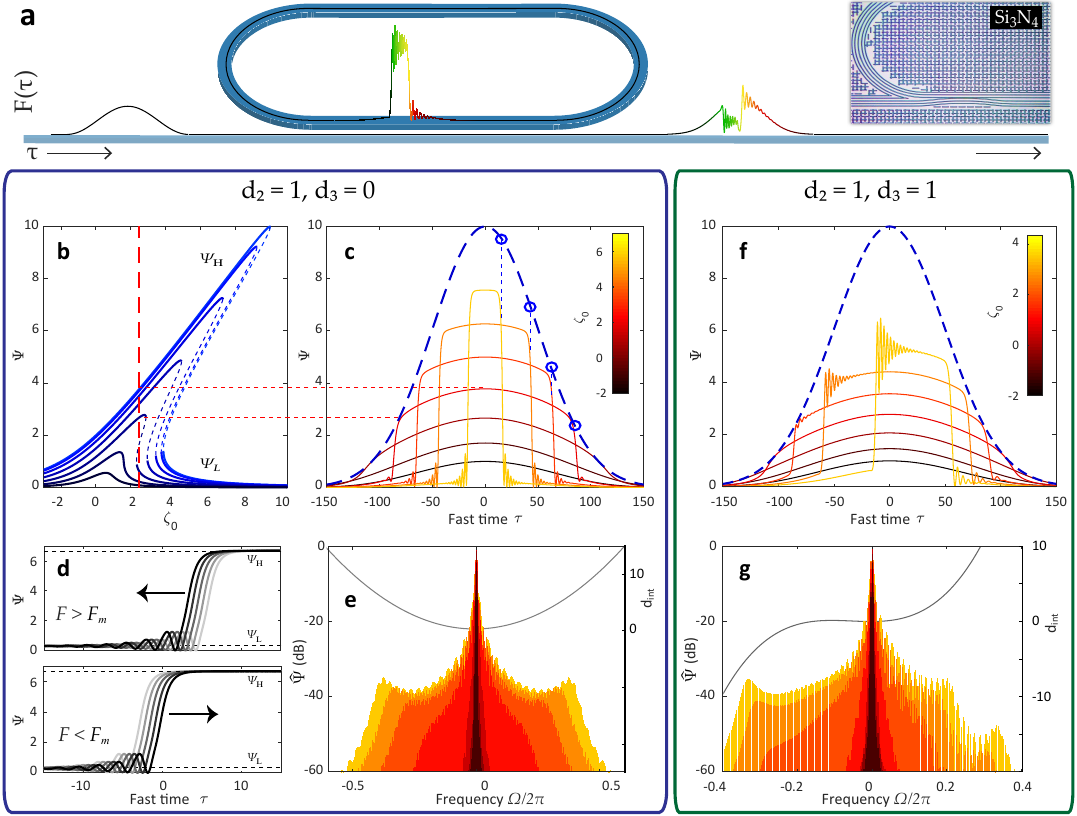}
	\caption{
		\textbf{Excitation of switching-waves inside pulse-drive envelope. }(a) Principle of pulse-driven Kerr cavity dissipative structure formation, with image of resonator with coupling section.  \textbf{(b-e) SW formation in pure normal dispersion. }(b)  Contour of the bistable intracavity CW solutions, plotted for increasing local value of the pump $F(\tau)$, with stable (unstable) solution in solid (dashed) line. (c) Development of intracavity field (red-yellow) within the pulse envelope (blue dashed) with increasing detuning $\zeta_0$. Red-dotted lines connect one field slice with the CW solution distribution in (b). Maxwell points on the pulse envelope marked with circles. (d) Expanding (top) and contracting (bottom) high-state (dashed) under CW driving. (e) Spectra of the fields from (c), and integrated dispersion operator. \textbf{(f-g) SW formation with strong third-order dispersion. }(f) Intracavity field within the pulse envelope with increasing detuning $\zeta_0$. (g) Spectra of the fields from (f).}
	\label{fig:concept2}
\end{figure*}

%%%%%%%%%%%%%%%%% Abstract

\textbf{Solitons are shape preserving waveforms that are ubiquitous across many nonlinear dynamical systems \cite{akhmediev_dissipative_2008}, and fall into two separate classes, that of bright solitons, formed in the anomalous group velocity dispersion regime, and `dark solitons'\cite{kivshar_dark_1998,godey_stability_2014} in the normal dispersion regime.  Both types of soliton have been observed in BEC \cite{becker_oscillations_2008}, hydrodynamics\cite{chabchoub_hydrodynamic_2016}, polaritons\cite{amo_polariton_2011}, and mode locked lasers\cite{grelu_dissipative_2012,liu_generation_2015}, but have been particularly relevant to the generation of microresonator-based frequency combs, where they have unlocked chipscale microcombs used in numerous system level applications in timing, spectroscopy, and communications\cite{kippenberg_dissipative_2018}.
For microcombs, both bright dissipative solitons, and alternatively dark pulses based on interlocking switching waves under normal dispersion, have been studied.
Yet, the existence of localized dissipative structures that fit between this dichotomy \cite{wai_soliton_1987} has been theoretically predicted \cite{parra-rivas_coexistence_2017}, but proven experimentally elusive.
Here we report the discovery of dissipative structures that embody a hybrid between switching waves and dissipative solitons, existing in the regime of (nearly) vanishing group velocity dispersion where third-order dispersion is dominant, hence termed as `zero-dispersion solitons'. These zero dispersion solitons are formed through collapsing perturbed switching wave fronts \cite{lobanov_generation_2019}, forming clusters of quantized solitonic sub-structures, which we synthesize in discrete numbers.
The switching waves are formed directly via synchronous pulse-driving\cite{obrzud_temporal_2017} of a photonic chip-based \SiN microresonator with vanishing normal dispersion. The resulting frequency comb spectrum is extremely broad in both the switching wave and zero-dispersion soliton regime, reaching 136 THz or 97\% of an octave. Fourth-order dispersion engineering results in dual-dispersive wave formation, and a novel quasi-phase matched dispersive wave related to Faraday instability\cite{copie_competing_2016}. This exotic, unanticipated dissipative structure expands the domain of Kerr cavity physics to the regime near to zero-dispersion, and could present a superior alternative to conventional bright solitons for broadband comb generation, but equally may find observation in other fields.}

%%%%%%%%%%%%%%%%%

\section[]{Introduction}

Currently, the field of research in optically driven Kerr nonlinear resonators and dissipative structure formation has been largely focused on the paradigm of the bright dissipative soliton \cite{kippenberg_dissipative_2018, herr_temporal_2014, leo_temporal_2010}. Bright dissipative solitons (DS) can be thought of as a particular variety of localized dissipative structure (LDS), solitary pulses that retain their shape due to the counter-balance between anomalous dispersion and nonlinearity, and who have a fixed amplitude determined by the driven-dissipative parameters of the Kerr cavity environment \cite{akhmediev_dissipative_2008}. Bright DS have been widely studied experimentally in multiple material platforms \cite{kippenberg_dissipative_2018}, and have been demonstrated as a desirable candidate for numerous integrated frequency comb-based applications such as massively parallel telecommunications \cite{marin-palomo_microresonator-based_2017} and LiDAR \cite{riemensberger_massively_2020}, astro-spectrometer calibration \cite{ewelina_obrzud_microphotonic_2019,suh_searching_2019}, dual-comb spectroscopy \cite{suh_microresonator_2016}, and also for metrology enabled by self-referencing such as absolute frequency synthesis \cite{spencer_optical-frequency_2018} and towards optical atomic clocks \cite{newman_architecture_2019}.
Across optical physics, DS have been observed in nonlinear systems such as mode-locked lasers, and transverse laser cavities \cite{lugiato_spatial_1987,ackemann_dissipative_2005,akhmediev_dissipative_2008}, and more widely as basic structures in nonlinear dynamical systems as diverse as plasma physics, neuron propagation, and chemical reaction systems \cite{nozaki_chaotic_1985, gonzalez-perez_solitary_2016, liehr2013dissipative}.

In opposition to bright DS have been dark dissipative structures, commonly called `dark pulses', which conversely exist in Kerr cavities possessing \emph{normal} dispersion \cite{godey_stability_2014,liang_generation_2014,xue_mode-locked_2015, huang_mode-locked_2015}. These dark pulses (alternatively termed as `platicons' \cite{lobanov_frequency_2015}) are in fact formed by the interlocking of two separate \emph{switching waves} (SW), connecting the high and low stable states of the bistable Kerr cavity \cite{parra-rivas_origin_2016}. Compared to bright dissipative solitons, they have been found to possess an intrinsically higher optical conversion efficiency between the input pump and the generated light as normal dispersion allows more comb lines far from the pump to be on resonance \cite{xue_microresonator_2017}. As such, they have been proposed as a superior alternative to bright DS for applications which require the generation of strong comb lines near to the pump center, and have been demonstrated as a source for massively parallel telecommunications \cite{fulop_high-order_2018}. Switching waves have been classed more generally as `domain walls' connecting two stable homogeneous states in driven dissipative systems, seen in $\upchi^{(2)}$ optical parametric oscillators \cite{rozanov_transverse_1982, trillo_stable_1997}, semiconductor lasers \cite{ganne_optical_2001}, birefringent optical fibers \cite{malomed_optical_1994, garbin_dissipative_2020}, and more widely in hydrodynamic systems and more \cite{pomeau_front_1986}.

In this work, we report the experimental discovery of a third LDS, a \emph{zero-dispersion} dissipative soliton (ZDS), which exists at the crossing point between conventional dissipative solitons and switching waves in the presence of vanishing second-order dispersion (SOD) giving way to pure- or dominant- third order dispersion (TOD).
It has remained an open question experimentally whether such a physical link exists. In the simplified case of pure SOD, these two cases of LDS are diametrically opposed. However, in realistic Kerr resonators, in the limit of vanishing SOD, there always exists a TOD component that begins to dominate. DS existing with a TOD component have been well demonstrated, but only in the sense where TOD acts as an additional perturbation to a regular DS, forming a dispersive wave tail that is far-removed from the center of the soliton spectrum \cite{jang_observation_2014,brasch_photonic_2016}. On the SW side, only previous theoretical work has investigated the case of strong TOD, and has predicted that in this regime, bright solitary pulses, amounting to a self-interlocked SW, may exist \cite{milian_soliton_2014, parra-rivas_coexistence_2017, bao_high-order_2017, lobanov_generation_2019}. In this sense, as we depict in Fig. \ref{fig:concept01}, it can be seen that there exists a continuous physical link between conventional DS and SW structures as a circular path of dispersion is followed in the SOD/TOD plane. In this depiction (based on numerical simulations), conventional DS become increasingly dominated by their dispersive wave tail, until when crossing into the region of normal dispersion, where individual zero-dispersion solitons are able to form packed clusters. When normal SOD begins to dominate once more, this cluster becomes unlinked into two separate switching waves perturbed by TOD, before reaching conventional pure-SOD switching waves.

In this work, we focus on the formation of ZDS on the normal SOD side of this plane. We find that this can be accomplished through first generating switching waves efficiently via \emph{wave-breaking} using synchronous pulse-driving of the Kerr cavity\cite{obrzud_temporal_2017} (Fig. \ref{fig:concept2}(a)), initially demonstrated in fiber cavities \cite{coen_convection_1999,luo_resonant_2015}, and proposed for microresonators \cite{lobanov_generation_2019}, and observing how these SWs coalesce into newly identified single ZDS states at high pump-cavity detuning.

\section{Theory}

To analyze the optical structures introduced in this work in a simple and universal fashion, we first consider an optical system described by the dimensionless Lugiato-Lefever Equation (LLE) \cite{coen_universal_2013, haelterman_dissipative_1992}, now with a non-CW driving term $f(\tau)$:

\begin{align}
	\label{eq:LLE1}
	\frac{\partial \psi}{\partial t'} = \left( -d_1\frac{\partial}{\partial\tau} -id_2\frac{\partial^2}{\partial\tau^2} +d_3\frac{\partial^3}{\partial\tau^3} \right)\psi & \nonumber \\ + ( i|\psi|^2 -i\zeta_0 -1 ) \psi + \sqrt{F_0}f(\tau) &
\end{align}

Here, the form taken by the field solutions are determined solely by the driving strength $F_0$ and detuning $\zeta_0$, as well as three parameters $d_l$ describing the relative contributions of the first three orders of dispersion \cite{milian_soliton_2014}. For simplicity, we set the SOD parameter $d_2=1$ throughout this work, which corresponds to normal dispersion. Thus, $d_3$ describes the contribution of TOD relative to $d_2$, and the first-order dispersion $d_1$ corresponds to the offset in group-velocity between the cavity field $\psi(\tau)$ and the static frame of the pulse-driving term $f(\tau)$.

Firstly, it is necessary to investigate direct SW formation by pulse-driving in the simplified case of pure SOD ($d_1=d_3=0$). We choose a value of $F_0=10$ which is a typical operating point for practical dissipative structure formation in experiment, and we set a Gaussian pulse as the driving function $f(\tau)=\exp(-\tau^2/\tau_p^2)$, with pulse duration $\tau_p=100$ so as to ensure any SW is significantly shorter in duration than the background driving function (which is true also in our experiment). The detuning is swept linearly from some value $\zeta_0<0$ up to $\zeta_0=10$. This range covers the region of Kerr cavity bistability, the CW solutions of which we graph in Fig. \ref{fig:concept2}(b) for different temporal samples of the pulse-drive amplitude $F(\tau)=F_0f^2(\tau)$, over $-120\leq\tau\leq0$. This effectively gives the \emph{local} CW solution of the intracavity field over the length of the pulse-drive envelope, divided into the high-state and low-state solutions $\psi_H$ and $\psi_L$, which can be solved analytically (see Methods).

In Fig. \ref{fig:concept2}(c) (with spectra in \ref{fig:concept2}(e)) we show the intracavity field solutions $\Psi=|\psi|^2$ at different values of $\zeta_0$ found using the split-step method \cite{coen_modeling_2013} (see Methods). For this direction in $\zeta_0$, the field initially follows the high-state solution $\psi_H(\tau)$ of the Kerr hysteresis. As $\zeta_0$ crosses 0, there begin to exist parts of the intracavity field where the local Kerr resonance-shift at the edges of the pulse-drive $F(\tau)$ is insufficient to sustain the high-state $\psi_H(\tau)$.
Here, the field outside this point falls to the low-state $\psi_L(\tau)$ while the field further inside the pulse background stays on $\psi_H(\tau)$ creating the SW that connects the two states \cite{rozanov_transverse_1982}.

\begin{figure*} [t]
	\includegraphics{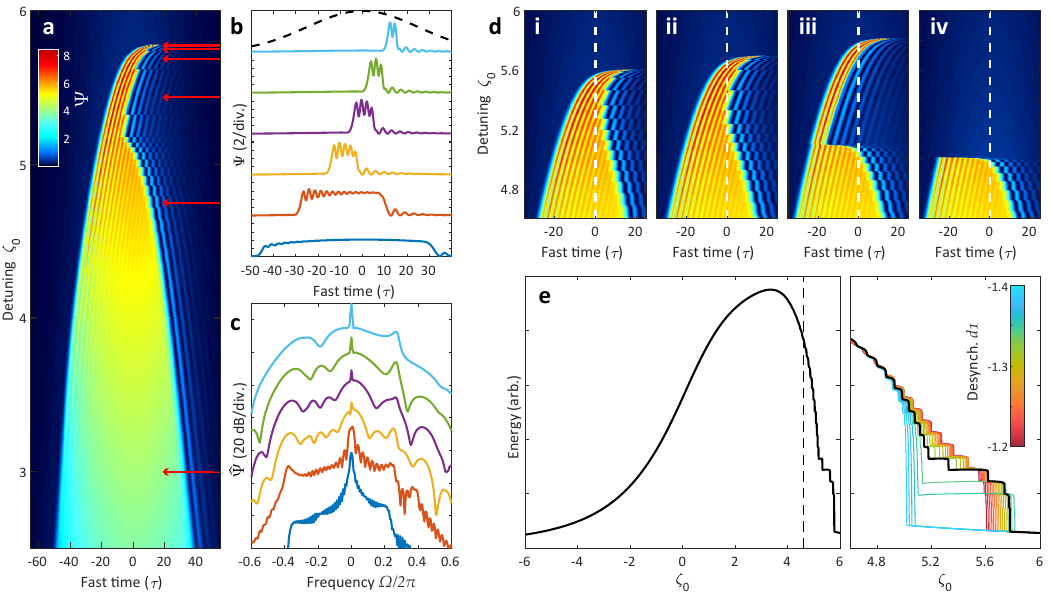}
	\caption{\textbf{Simulation: zero-dispersion soliton formation via desynchronized pulse-driving. $F_0=10, d_2=1, d_3=1$ }(a) Intracavity field as detuning $\zeta_0$ is increased. $d_1=-1.34$ (b) Individual time-domain and (c) frequency-domain of the fields in (a) (red arrows). (d) Alternative field formation with (i) $d_1=-1.20$ (ii) $d_1=-1.29$ (iii) $d_1=-1.36$ (iv) $d_1=-1.37$. (e) (left) Total intracavity energy from (a) (right) Zoom of the `step' feature, plotted for different values of desynchronization $d_1$. The black line corresponds to the field given in (a).}
	\label{fig:conceptTOD}
\end{figure*}

From here, the two SW locations $\tau_\mathrm{SW}$ follow a location within the pulse-drive envelope $F(\tau)=F_m$, which previous theoretical works on SW stability have termed as the `Maxwell Point' \cite{ganne_optical_2001, parra-rivas_origin_2016}, until at $\zeta_0\approx7$ where there exists no $F(\tau)>F_m$ causing the SWs to meet each other and annihilate, failing to reach their theoretical maximum detuning at $\zeta_0=F_0=10$.
The stability of the SW fronts within the pulse envelope after formation is due to the  effective `outward pressure' manifesting on $\psi_H$. SWs possess an innate group velocity offset depending on the value of $F$ and $\zeta_0$ \cite{coen_convection_1999}, where the $\psi_H$ tends to undergo expansion, with the SWs moving outward, when the driving term is larger than a certain value $F(\tau)>F_m$ for a fixed detuning $\zeta_0$ (see Fig. \ref{fig:concept2}(d)). When $F(\tau)<F_m$, the $\psi_H$ contracts and the SWs move inward. Accepting this, it becomes clear that if any high-state $\psi_H$ existed within a pulse-drive envelope, whose peak $F_0>F_m$, it would undergo expansion until its SW fronts reached a point where $F(\tau_\mathrm{SW})=F_m$ and stop.

Considering now a Kerr cavity possessing strong TOD, we choose $d_3=1$. Fig. \ref{fig:concept2}(f,g) presents an identical scenario to Fig. \ref{fig:concept2}(b-e), now with TOD enabled. Here, time-reversible symmetry has been broken, with a strong difference emerging between the up-SW (here left in time), and the down-SW (on the right). The down-SW, which corresponds with the dispersive wave-like feature on the positive side of the spectrum (Fig. \ref{fig:concept2}(g)), follows similar behavior to that of the SWs under pure SOD in Fig. \ref{fig:concept2}(c,e) including the Maxwell point as detuning $\zeta_0$ is increased. Conversely, the up-SW exists primarily on the negative side of the spectrum, where the local gradient of the integrated dispersion operator $d_\mathrm{int}=d_2\Omega^2+d_3\Omega^3$ ($\Omega$ being the dimensionless angular frequency) has become inverted. Consequently, the dispersive wave element now possesses a positive group-delay (negative group-velocity shift) and so has changed orientation to face inward to the high-state $\psi_H$, and the up-SW itself has had its Maxwell point changed so that it retreats inside the pulse-drive envelope much sooner, thus eliminating the SW fronts at a much earlier detuning at $\zeta_0\approx4.5$.

Overall, the flatness of the dispersion profile on negative frequencies has heavily skewed the generated spectrum to the one side resulting in a negative shift to the group velocity for the SW structure as a whole inside $F(\tau)$. Naturally, introducing a counter-acting group velocity shift in the form of a negative $d_1$ term should help contain the structure within the center of $F(\tau)$ as the detuning increases. This scenario is presented in Fig. \ref{fig:conceptTOD}. By now setting $d_1=-1.34$, the time-frame of the cavity field continually moves forward in fast time $\tau$ (here to the left), keeping both SWs near to the center of the pulse envelope $F(\tau)$ preventing early collapse. The SW fronts meet together now at $\zeta_0=5.1$, where a significant event occurs. Instead of eliminating each other as in the case of pure-SOD, the SWs become locked to each other based on the bonding of the down-SW to the modulated wave of the up-SW, in an event not at all dissimilar to the much reported formation of `dark pulses' \cite{xue_mode-locked_2015}.  Whereas dark pulses lock on the modulations of the low-state $\psi_L$, these `bright' structures necessarily require strong TOD so that a sufficiently powerful modulation exists on the high-state $\psi_H$.

These bright LDS, whose existence was predicted in recent theoretical works \cite{parra-rivas_coexistence_2017, lobanov_generation_2019} have been termed as highly modulated `platicons'. However, like dissipative solitons, these structures are self-stable and can freely exist across the broad background of the pulse-drive.  While the SW states shown here for $\zeta_0<5.1$ are bound to the pulse-drive envelope $F(\tau)$ as a whole through the left and right Maxwell points, these bright structures are self-stable and eventually find a single \emph{trapping position} on one edge of $F(\tau)$ just as with pulse-driven conventional dissipative solitons \cite{obrzud_temporal_2017, hendry_impact_2019-1, anderson_photonic_2019}. In this example, the trapping position is on the left edge.  As the detuning here increases past $\zeta_0>5.1$, the structure (plotted specifically on levels 3-6 of Fig. \ref{fig:conceptTOD}(b,c)) undergoes progressive collapses reducing in periodicity initially from 5, down to 2. In this way, these structures can be thought of as clusters of individual dissipative solitons packed extremely close together. As their existence is defined by vanishing SOD in favor of TOD, we term them in this work as zero-dispersion solitons (ZDS$^{(n)}$) consisting of $n$ bound sub-solitons. The value $n$ can be identified by the spectral periodicity between the pump and what was initially the SW dispersive wave, here on the left end of the spectrum.

Fig. \ref{fig:conceptTOD}(d) shows how varying the group velocity shift $d_1$ (or \emph{desynchronization} in terms of pulse-driving) gives rise to a varying maximum detuning for ZDS$^{(n)}$ existence, and with different preferred $n$.
Here (with particular attention to Fig. \ref{fig:conceptTOD}(d-iii)), the ZDS$^{(3)}$ follows its trapping position on the left-hand slope until it crosses the center line where a trapping position no longer exists and decays, following an asymmetrical trajectory reminiscent of recent studies on conventional dissipative solitons \cite{hendry_impact_2019-1}. The cavity energy trace, plotted in Fig. \ref{fig:conceptTOD}(e) for all values of $d_1$ in the vicinity, shows the asymmetrical unfolding of the characteristic `step' feature we should expect to see in experiment.

\begin{figure*} [t]
	\includegraphics[width=0.95\textwidth]{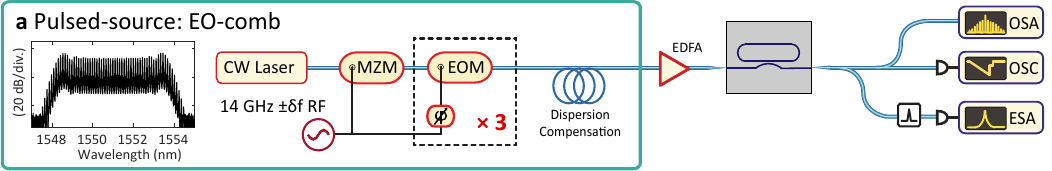}
	\includegraphics[width=0.95\textwidth]{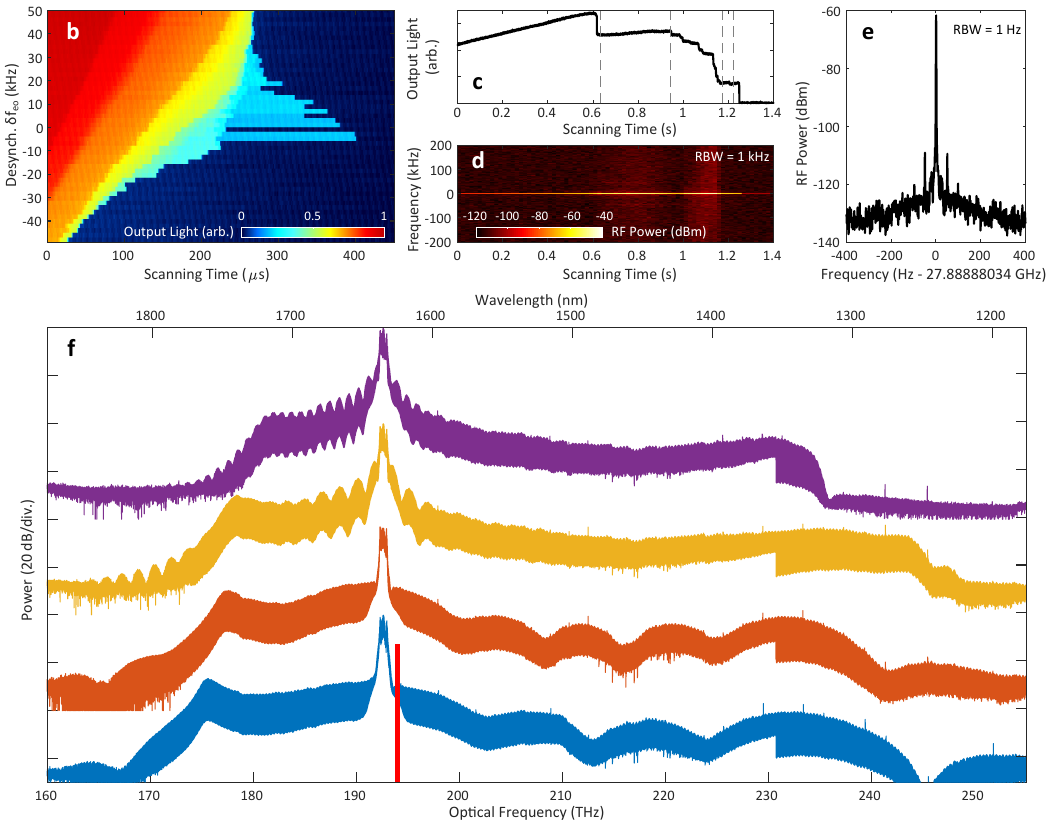}
	\caption{\textbf{Experimental pulse-driven switching wave and zero-dispersion soliton formation, MR1} (a) Setup, featuring the EO-comb as a pulsed-source.  MZM: Mach-Zehnder modulator, EOM: electro-optic modulator, EDFA: erbium-doped fiber amplifier, ESA: electronic spectrum analyzer, OSA: optical spectrum analyzer, OSC: oscilloscope. The input pulse train is coupled into and out of the microresonator chip via lensed fibers. Left-inset: Spectrum of the 14 GHz EO-comb before amplification.  (b) Spectrogram of the step feature for different desynchronization about 27.888880 GHz.  (c) Microresonator transmission (with DC value subtracted), with detunings from (f) marked with dashed lines.  (d) Spectrogram of the repetition-rate beatnote during the laser scan in (c).  (e) Long-term beatnote measurement of the final comb state. (f) Stages of comb/spectrum formation in descending order of detuning (40 dB vertical offsets). Red block marks spectral filter for beatnote measurement. }
	\label{fig:experiment1}
\end{figure*}

\begin{figure*} [t]
	\includegraphics{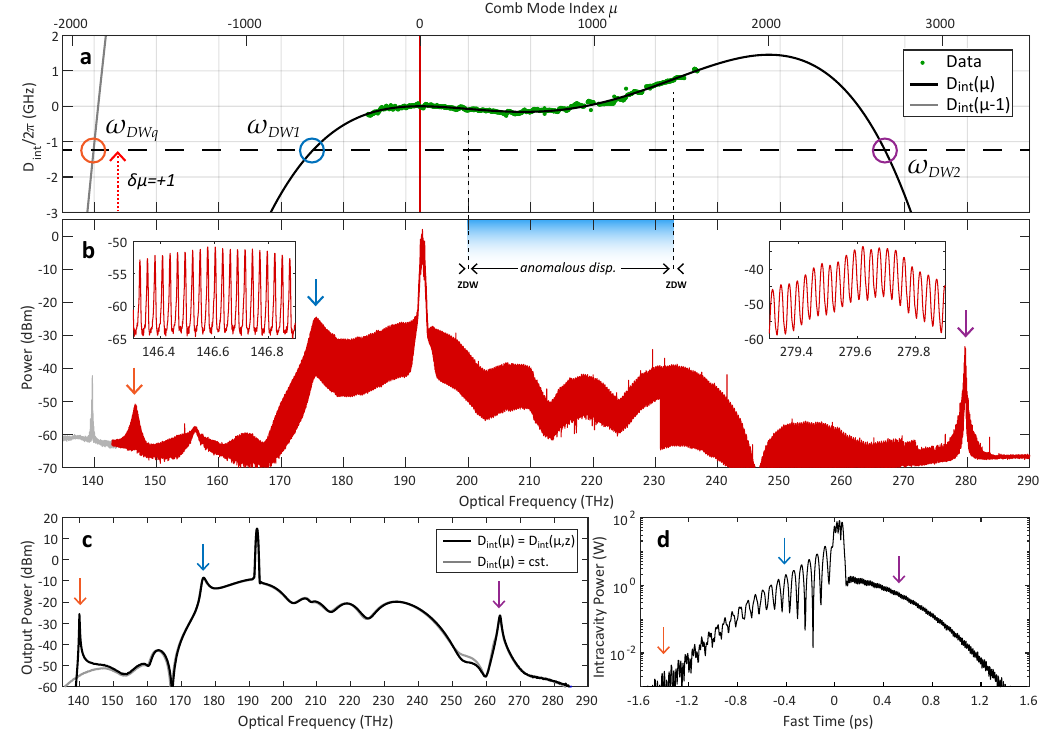}
	\caption{\textbf{Octave spanning zero-dispersion soliton spectrum, MR1}(a) Experimental measurement of the resonator integrated dispersion $D_\mathrm{int}$, along with the spectrally extended fitted solution, and solution shifted by $+D_1$. Pump frequency (and pump detuning) marked by the vertical red line (horizontal dashed line). Phase-matched locations of dispersive waves marked by circles, with momentum mismatch in dotted red arrow.  (b) Measured spectrum of ZDS$^{(4)}$, with dispersive waves marked with arrows corresponding to the circles in (a). Insets show individual comb lines. The left-most trace is in gray to indicate it is the second-order diffraction spectrum of DW2, and not genuine.  (c) Frequency-domain and (d) time-domain simulation of ZDS$^{(4)}$ with dispersive wave tails marked.}
	\label{fig:experiment2}
\end{figure*}

\begin{figure} [h]
	\centering
	\includegraphics{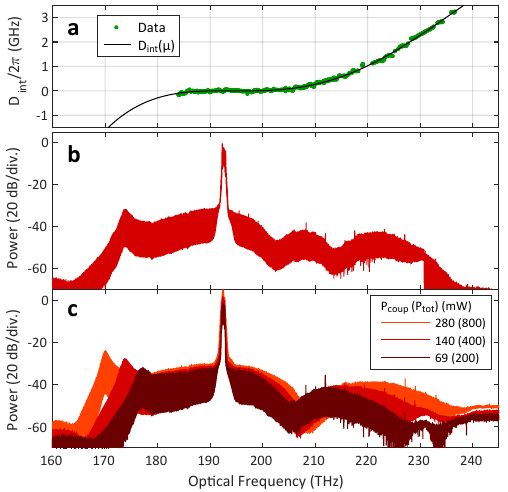}
	\caption{\textbf{Experimental zero-dispersion soliton formation, MR2} (a) Experimental measurement of the resonator integrated dispersion $D_\mathrm{int}$, and spectrally extended fitted solution.  (b) ZDS$^{(3)}$ comb.  (c) ZDS$^{(2)}$ combs, measured at maximum detuning using three pump powers, noted as effective average power coupled to ring (total power towards chip). }
	\label{fig:experiment4}
\end{figure}

\section{Experimental Results}
The pulse-drive source (as shown in Fig. \ref{fig:experiment1}(a)) is provided in the form of an electro-optic comb (EO-comb) \cite{kobayashi_optical_1988,obrzud_temporal_2017}, providing pulses with a minimum duration of 1 ps (see Methods for details), and whose repetition rate is finely controlled by an RF-synthesized signal $f_\mathrm{eo}$. The cavity platform of choice for the experiment is the chip-based \SiN microresonator, in this case having a native FSR of 27.88 GHz. The EO-comb repetition rate is set to exactly half this at $f_\mathrm{eo}=13.944$ GHz due to RF transmission limitations. Other than a factor-2 reduction on conversion efficiency due to only half the lines being coupled to the cavity, the experiment behaves the same as one that is fully synchronous and we can disregard the excess comb lines. Two microresonators (referred to as MR1 and MR2) in particular are used to generate ZDS, having two slightly different dispersion profiles causing the formation of ZDS$^{(n)}$ of different $n$. Their measured dispersion parameters are given in result figures further below.

Starting with MR1, in order to ensure that a full range of formation behavior is observed, and spectral extent maximized, the average pulse-power coupled to the resonator is set to $P_\mathrm{av}=350$ mW (12 pJ pulse energy at 28 GHz), approximately 20 times higher than the observed minimum comb generation threshold. The experiment proceeds the same way as in the theory section, typical for LDS generation in Kerr cavities, and particularly in pulse-driven soliton generation \cite{obrzud_temporal_2017, lilienfein_temporal_2019, anderson_photonic_2019}. The exact native FSR of the microresonator is first found by varying the input repetition rate $f_\mathrm{eo}$ until the expected unfolding of the ZDS `step' is observed (Fig. \ref{fig:experiment1}(b)). Here, we see an asymmetrical extension of the step vs. the relative desynchronization $\delta\!f_\mathrm{eo}=f_\mathrm{eo}-D_1/2\pi$, as expected based on Fig. \ref{fig:conceptTOD}(e), although with slightly different precise form due to unaccounted for higher order effects (the absence of a step at $\delta\! f_\mathrm{eo}=0$ is a coincidence based on shot-to-shot statistical variation of formation probability). Based on this measurement, we find an optimum $f_\mathrm{eo}=27.88888$ GHz, with a locking range for ZDS on the order of $\pm10$ kHz.

For this value of $f_\mathrm{eo}$, the EO-comb seed laser frequency $\omega_p$ is tuned slowly across a resonance frequency $\omega_0$ from the blue-detuned side to the red-detuned side (such that $\delta\omega=\omega_0-\omega_p>0$ by convention\cite{herr_temporal_2014}) towards the region of Kerr bistability. Fig. \ref{fig:experiment1}(c) plots the output light from the microresonator during this scan, and at the same time the RF repetition-rate beatnote of the ZDS is recorded (Fig. \ref{fig:experiment1}(d)). The measured comb spectra are plotted in Fig. \ref{fig:experiment1}(f) at four detunings in descending order, after the SW is formed. Qualitatively, the results behave as the simulations in Fig. \ref{fig:conceptTOD}(c) predict. Firstly, over the first two rows, we see the spectrum grow wider and with a sharper dispersive-wave (DW1) located from 182 to 179 THz. Importantly, we see the spectral fringes either side of the pump (spaced by $\approx1$ THz on the first spectrum) increase their period as detuning is increased, indicating the two SW fronts are moving together within the pulse-drive envelope. In the last two rows, we see the SWs have coalesced into the ZDS$^{(5)}$, a 5-period structure, then reducing to a ZDS$^{(4)}$, each time moving the location of DW1 further to low frequencies. Throughout this scan over resonance, the RF repetition rate beatnote shows low noise. In fact, the final beatnote, plotted in Fig. \ref{fig:experiment1}(e) is highly stable, inheriting the low-offset phase noise of the $f_\mathrm{eo}$ as supplied by the RF synthesizer. This confirms that the ZDS has temporally locked to the driving pulse, just as a conventional bright dissipative soliton would \cite{obrzud_temporal_2017, anderson_photonic_2019}.

Fig. \ref{fig:experiment2} analyses this final structure in greater detail. We characterize the broadband dispersion profile of the MR1 using a cascaded three-laser swept spectroscopy technique \cite{liu_frequency-comb-assisted_2016}. In Fig. \ref{fig:experiment2}(a) we plot the measured integrated dispersion profile $D_\mathrm{int}=\omega_\mu-\omega_0-\mu D_1$, representing the frequency deviation of each resonator mode $\mu$ from the uniform FSR grid spaced by $D_1$ (where the pump mode corresponds to $\mu=0$). This data is fitted to a fourth-order polynomial centered at $\omega_0=2\pi\times192.3$ THz where $D_\mathrm{int}\approx \mu^2D_2/2 +\mu^3D_3/6 +\mu^4D_4/24$ ($D_2/2\pi = -3.17$ kHz, $D_3/2\pi = 13.8$ Hz, $D_4/2\pi = -15.9$ mHz, all $\pm5\%$). In dimensionless parameters, for a fixed $d_2=1$ we have a value $d_3=0.38$. The pump frequency detuning $-\delta\omega/2\pi=-1.2$ GHz, which we obtain from live cavity phase-response measurements \cite{guo_universal_2017}, is also marked. In Fig. \ref{fig:experiment2}(b) the entire spectrum of the ZDS$^{(4)}$ is plotted and features several dispersive waves (DW), the spectral locations of which can be predicted based on where $D_\mathrm{int}(\mu)=-\delta\omega$. The predicted DW locations do not match perfectly with experiment however, but this can be explained by the bandwidth-limited dispersion measurement with unknown higher-order values for $D_5, D_6$ and so forth.

The first, DW1 at 176 THz, will always occur for ZDS formation due to the requirement for powerful TOD. The second, DW2 at 280 THz, has occurred due to the overall normal fourth-order dispersion of the waveguide \cite{pfeiffer_octave-spanning_2017}, but is not required for ZDS formation. The additional dispersive wave, termed DWq, occurs where the optical comb modes have wrapped by $D_1$ so that $D_\mathrm{int}(\mu)+D_1=D_\mathrm{int}(\mu-1)=-\delta\omega$ or that, by shifting by one FSR, the linear wave at DWq has accrued a $2\pi$ phase shift relative to the pump wave. This phase-wrapping is commonly known to allow the formation of so called `Kelly' sidebands in soliton fiber lasers \cite{kelly_characteristic_1992}. Due to the longitudinal momentum mismatch $\delta\mu=+1$ between the coupled linear wave and the ZDS comb lines, quasi-phase matching is required to bridge this gap \cite{hickstein_quasi-phase-matched_2018}. This particular microresonator features a brief `mode-stripping' section where the waveguide width rapidly tapers down to a narrow width in order to stop any higher-order spatial modes from propagating \cite{kordts_higher_2016}, where the waveguide dispersion changes sharply. This intra-roundtrip disturbance provides phase modulation to the linear wave at DWq, and is more than sufficient to enable quasi-phase matching to stimulate resonant radiation \cite{copie_competing_2016}. This effect has been observed in microresonators with a similar intra-roundtrip modulation of the waveguide width where it was identified as Faraday instability \cite{huang_quasi-phase-matched_2017}, and has long been observed in fiber-based Kerr resonators with longitudinally varying dispersion \cite{luo_resonant_2015, copie_competing_2016, nielsen_invited_2018}.
Fig. \ref{fig:experiment2}(c,d) shows numerical LLE simulations using the real experimental parameters of MR1 (see Methods), demonstrating close agreement with the form taken by the spectrum corresponding to a ZDS$^{(4)}$ as shown in Fig. \ref{fig:experiment2}(d). In Fig. \ref{fig:experiment2}(c), both simulation results taking into account either a constant or an oscillating intra-roundtrip dispersion $D_\mathrm{int}(z)$, with DWq appearing only in the latter case.  Further simulations and analysis DWq is presented in the supplementary info.

In order to observe ZDS$^{(n)}$ of lower $n$ we move to MR2, which has its zero-dispersion wavelength closer to the pump wavelength at 1560 nm. Here, the dispersion parameters (Fig. \ref{fig:experiment4}(a)) are fitted to be $D_2/2\pi = -848$ Hz, $D_3/2\pi = 12.8$ Hz, $D_4/2\pi = -15.9$ mHz, $\pm5\%$, corresponding to dimensionless parameters $d_2=1$ and $d_3=2.11$, further into the zero-dispersion regime (Fig. \ref{fig:concept01}). In this microresonator, with the same generation method as above in MR1, we generate ZDS$^{(3)}$ and ZDS$^{(2)}$ in Fig. \ref{fig:experiment4}(a) and (b) respectively. In this microresonator, we do not observe the same DW2 or DWq as in microresonator 1. For ZDS$^{(2)}$, we present spectral measurements taken using three increasing input pump powers, each enabling an increased maximum detuning $\delta\omega$. As shown, as available power is increased, the overall spectral profile expands, with both DW1 on the left and the anti-dispersive wave on the right moving outwards, in a similar manor as for conventional dissipative solitons \cite{lucas_detuning-dependent_2017}.

\section{Discussion and Summary}

In this work, we have experimentally synthesized a novel class of localized dissipative structure, the zero-dispersion soliton. In terms of figures of merit, the generated ZDS$^{(2-5)}$-based combs presented here are extremely substantial in terms of the product of their total bandwidth and their total line-count which, as far as we are aware, is a record for a single-structure in a microresonator. The central body of the ZDS$^{(4)}$ comb in MR1 spans over 76 THz (1830 and 1260 nm), accounting for more than 2,700 comb teeth, spaced by a detectable 28 GHz repetition rate. When including the DW features, the final bandwidth becomes 136 THz or 97\% of an octave. As the repetition rate is directly detectable on photodiode, a future work with fine-tuning of the microresonator dispersion may enable $f$-$2f$ self-referencing with a single microcomb \cite{spencer_optical-frequency_2018}. We have further demonstrated the \emph{direct} generation of switching waves via pulse-driving, creating a highly smooth ultra-broadband microcomb under normal dispersion conditions. Such normal dispersion-based microcombs have thus far only been formed in \SiN via modulation instability enabled by spatial mode-coupling\cite{xue_mode-locked_2015}, necessitating an extra coupled-microresonator ring with integrated heaters in order to be deterministic \cite{kim_turn-key_2019}.

The formation of ZDS-based microcombs more generally has expanded the domain of microcomb generation towards the region of both normal-dispersion and zero-dispersion, previously not often considered ideal. This lifting of strict requirements for anomalous dispersion may give greater flexibility in the \SiN fabrication process going forward. The result also demonstrates not only that microcomb generation can be achieved in a straight-forward fashion in such waveguide resonators with normal-to-zero dispersion, but that it may be most \emph{preferable} for highly broadband comb generation due to the superior flatness of the comb in the SW regime, as well as the lack of a high-noise chaotic phase and multi-soliton formation as compared to its anomalous dispersion-based counterpart. In terms of physics, the experimental observations of an entirely novel class of LDS -- a bright pulse-like structure which constitutes a link between SW-based and soliton-based LDS -- were presented. To our knowledge, such an entity has not been observed before in microresonators as of the time of writing. Its discovery in experiment may open a new area of fundamental research on the nature of dissipative Kerr solitons and switching waves under one umbrella.

Note: we would like to acknowledge a parallel work by Li \emph{et al.} completed during preparation for this manuscript observing a similar phenomenon in fiber-based Kerr cavities \cite{li_experimental_2020}.

%%%%%%%%%%%%%%%%%%%%%%%%%%%%%%%%%%%%%%%%%%%%

\subsection*{Acknowledgments}

This work was supported by Contract No. D18AC00032 (DRINQS) from the Defense Advanced Research Projects Agency (DARPA), and funding from the Swiss National Science Foundation under grant agreements No. 192293. This material is based upon work supported by the Air Force Office of Scientific Research, Air Force Materiel Command, USAF under Award No. FA9550-19-1-0250.

\subsection*{Author Contributions}

M.H.A. performed the experiment with the assistance of G. L.. The sample microresonator was fabricated by J.L.. M.H.A. conducted the simulations and prepared the manuscript with the assistance W.W., G.L., and T.J.K.. T.J.K. supervised the project.

\subsection*{Data Availability Statement}

All data and analysis files will be made available via \texttt{zenodo.org} upon publication.

\subsection*{Methods}

\begin{small}
	
\textbf{Theory and Simulation} 
The homogeneous solutions to the intracavity field $\Psi_\mathrm{CW}=|\psi_\mathrm{CW}|^2$, graphed in Fig. \ref{fig:concept2}(b) for different local pump strength $F(\tau)=F_0f^2(\tau)$, is first obtained from the real roots of the cubic polynomial derived from equation (\ref{eq:LLE1}) at equilibrium and with all dispersion $d_l=0$ \cite{haelterman_dissipative_1992}, and subsequently the complex field solution from the resonance condition,

\begin{equation}
	\Psi^3-2\zeta_0\Psi^2+(\zeta_0^2+1)\Psi-F(\tau)=0
\end{equation} 

\begin{equation}
\psi_\mathrm{CW}=\frac{i\sqrt{F(\tau)}}{\Psi_\mathrm{CW}-\zeta_0+i}
\end{equation} 

with $\Psi_H$ and $\Psi_L$ being the top and bottom solution respectively. The simulation presented in Fig. \ref{fig:concept2} was calculated via the split step method with a change in detuning rate $d\zeta_0/dt'=0.01$ to ensure SWs reached equilibrium at each stage. In Fig. \ref{fig:conceptTOD}, $d\zeta_0/dt'=0.000625$ to allow the ZDS to remain at their trapping/equilibrium position during he detuning increase. The pulse-drive width $\tau_p=50$.

\textbf{Experiment} 
The EO-comb is comprised of a CW laser, followed by an intensity modulator and three phase modulators, driven by an RF signal generator (Rhode \& Schwarz SMB100A), generating 50 spectral lines spaced by $f_\mathrm{eo}=$ 13.944 GHz. The waveform is compressed in time through linear dispersion made from 300 m of standard SMF-28 and 5 m of dispersion-compensating fiber, yielding pulses of minimum duration 1 ps as confirmed by frequency-resolved optical gating (FROG). 
The \SiN microresonators MR1 and MR2 used in this experiment have been fabricated with the \emph{photonic Damascene process} \cite{pfeiffer_photonic_2018} with a 2350$\times$770 nm$^2$ cross-section, and possess a peak probable cavity linewidth of $\kappa/2\pi=(\kappa_0+\kappa_\mathrm{ex})/2\pi=$ 208 MHz and 150 MHz (with external coupling rate $\kappa_\mathrm{ex}/2\pi=$ 155 MHz and 120 MHz) for MR1 and MR2 respectively. Their measured dispersion $D_\mathrm{int}$ is expanded in the main text.
Effective power coupled to resonator as quoted above exclude chip insertion loss of 1.6 dB and half of the 14 GHz comb lines not coupled to the resonator modes at 28 GHz. The RF beatnote measurement in Fig. \ref{fig:experiment1}(d,e) derives from approximately 11 filtered comb lines outside of the EO-comb spectrum.

\textbf{Full system model}
The experimental microresonator results are described by the full LLE with real parameters
%$$
\begin{align}
%	\label{eq:LLE2}
	\frac{\partial A(t,T)}{\partial t} =\quad & \mathcal{F} \Big[ i\big( \delta\omega +\mu \cdot2\pi\delta\!f_\mathrm{eo} +D_\mathrm{int}(z,\mu) \big) \tilde{A}_\mu \Big] \\ \nonumber - & \frac{\kappa}{2}A  + ig_0|A|^2A +\sqrt{\frac{\kappa_\mathrm{ex}P_0}{\hbar\omega_0}} f_p(T)
\end{align}
%$$	
acting on photon field $A(t,T)$ over slow/laboratory time $t$ and fast time $T$ in the co-moving frame of the intracavity field circulating at $D_1$, with frequency domain counterpart $\tilde{A}_\mu$ at discrete comb line indices $\mu$. Included with the linear phase operators $D_\mathrm{int}$ and $\delta\omega$ is the input pulse desynchronization $\delta\!f_\mathrm{eo}$. The nonlinear coupling parameter $g_0/2\pi=0.056$ Hz (see supp. info for further on this). 

For the simulation presented in Fig. \ref{fig:experiment2}(c,d), we set $\delta\omega/2\pi=$ 900 MHz, and the input pulse profile $f_p(T)=\exp(-T^2/T_p^2)$ with $T_p=$ 0.85 ps, $P_0=1.8$ W, and static desynchronization $\delta\!f_\mathrm{eo}=150$ kHz.  In order to stimulate the quasi phase-matched wave at DWq, we set $D_\mathrm{int}(z,\mu)=D_\mathrm{int0}(\mu)(1+0.2\cos(\delta\mu \cdot2\pi z/L))$, with $\delta\mu=1$ representing a single longitudinal-mode modulation in the dispersion operator for the resonator of length $L$. All of the real parameters are related to the dimensionless parameters by the following: $t'=\frac{\kappa}{2}t$, \quad$\tau=D_1\sqrt{\frac{\kappa}{D_2}}T$, \quad$\psi=\sqrt{\frac{2g_0}{\kappa}}A$, \quad$d_l=\frac{2}{\kappa}\frac{D_l}{l!} \big( \frac{\kappa}{D_2} \big)^\frac{l}{2}$ for $l=$1-4, \quad$\zeta_0=\frac{2\delta\omega}{\kappa}$, \quad$F_0=\frac{8\kappa_\mathrm{ex}g_0}{\kappa^3\hbar\omega_0}P_0$.

\end{small}

%%%%%%%%%%%%%%%%%%%%%%%%%%%%%%%%%%%%%
\bibliographystyle{apsrev4-2}
\bibliography{biblib_2}

\begin{thebibliography}{66}%
\makeatletter
\providecommand \@ifxundefined [1]{%
 \@ifx{#1\undefined}
}%
\providecommand \@ifnum [1]{%
 \ifnum #1\expandafter \@firstoftwo
 \else \expandafter \@secondoftwo
 \fi
}%
\providecommand \@ifx [1]{%
 \ifx #1\expandafter \@firstoftwo
 \else \expandafter \@secondoftwo
 \fi
}%
\providecommand \natexlab [1]{#1}%
\providecommand \enquote  [1]{``#1''}%
\providecommand \bibnamefont  [1]{#1}%
\providecommand \bibfnamefont [1]{#1}%
\providecommand \citenamefont [1]{#1}%
\providecommand \href@noop [0]{\@secondoftwo}%
\providecommand \href [0]{\begingroup \@sanitize@url \@href}%
\providecommand \@href[1]{\@@startlink{#1}\@@href}%
\providecommand \@@href[1]{\endgroup#1\@@endlink}%
\providecommand \@sanitize@url [0]{\catcode `\\12\catcode `\$12\catcode
  `\&12\catcode `\#12\catcode `\^12\catcode `\_12\catcode `\%12\relax}%
\providecommand \@@startlink[1]{}%
\providecommand \@@endlink[0]{}%
\providecommand \url  [0]{\begingroup\@sanitize@url \@url }%
\providecommand \@url [1]{\endgroup\@href {#1}{\urlprefix }}%
\providecommand \urlprefix  [0]{URL }%
\providecommand \Eprint [0]{\href }%
\providecommand \doibase [0]{https://doi.org/}%
\providecommand \selectlanguage [0]{\@gobble}%
\providecommand \bibinfo  [0]{\@secondoftwo}%
\providecommand \bibfield  [0]{\@secondoftwo}%
\providecommand \translation [1]{[#1]}%
\providecommand \BibitemOpen [0]{}%
\providecommand \bibitemStop [0]{}%
\providecommand \bibitemNoStop [0]{.\EOS\space}%
\providecommand \EOS [0]{\spacefactor3000\relax}%
\providecommand \BibitemShut  [1]{\csname bibitem#1\endcsname}%
\let\auto@bib@innerbib\@empty
%</preamble>
\bibitem [{\citenamefont {Akhmediev}\ and\ \citenamefont
  {Ankiewicz}(2008)}]{akhmediev_dissipative_2008}%
  \BibitemOpen
  \bibinfo {editor} {\bibfnamefont {N.}~\bibnamefont {Akhmediev}}\ and\
  \bibinfo {editor} {\bibfnamefont {A.}~\bibnamefont {Ankiewicz}},\ eds.,\
  \href {https://www.springer.com/gp/book/9783540782162} {\emph {\bibinfo
  {title} {Dissipative {Solitons}: {From} {Optics} to {Biology} and
  {Medicine}}}},\ Lecture {Notes} in {Physics}\ (\bibinfo  {publisher}
  {Springer-Verlag},\ \bibinfo {address} {Berlin Heidelberg},\ \bibinfo {year}
  {2008})\BibitemShut {NoStop}%
\bibitem [{\citenamefont {Kivshar}\ and\ \citenamefont
  {Luther-Davies}(1998)}]{kivshar_dark_1998}%
  \BibitemOpen
  \bibfield  {author} {\bibinfo {author} {\bibfnamefont {Y.~S.}\ \bibnamefont
  {Kivshar}}\ and\ \bibinfo {author} {\bibfnamefont {B.}~\bibnamefont
  {Luther-Davies}},\ }\href {https://doi.org/10.1016/S0370-1573(97)00073-2}
  {\bibfield  {journal} {\bibinfo  {journal} {Physics Reports}\ }\textbf
  {\bibinfo {volume} {298}},\ \bibinfo {pages} {81} (\bibinfo {year}
  {1998})}\BibitemShut {NoStop}%
\bibitem [{\citenamefont {Godey}\ \emph {et~al.}(2014)\citenamefont {Godey},
  \citenamefont {Balakireva}, \citenamefont {Coillet},\ and\ \citenamefont
  {Chembo}}]{godey_stability_2014}%
  \BibitemOpen
  \bibfield  {author} {\bibinfo {author} {\bibfnamefont {C.}~\bibnamefont
  {Godey}}, \bibinfo {author} {\bibfnamefont {I.~V.}\ \bibnamefont
  {Balakireva}}, \bibinfo {author} {\bibfnamefont {A.}~\bibnamefont
  {Coillet}},\ and\ \bibinfo {author} {\bibfnamefont {Y.~K.}\ \bibnamefont
  {Chembo}},\ }\href {https://doi.org/10.1103/PhysRevA.89.063814} {\bibfield
  {journal} {\bibinfo  {journal} {Physical Review A}\ }\textbf {\bibinfo
  {volume} {89}},\ \bibinfo {pages} {063814} (\bibinfo {year}
  {2014})}\BibitemShut {NoStop}%
\bibitem [{\citenamefont {Becker}\ \emph {et~al.}(2008)\citenamefont {Becker},
  \citenamefont {Stellmer}, \citenamefont {Soltan-Panahi}, \citenamefont
  {Dörscher}, \citenamefont {Baumert}, \citenamefont {Richter}, \citenamefont
  {Kronjäger}, \citenamefont {Bongs},\ and\ \citenamefont
  {Sengstock}}]{becker_oscillations_2008}%
  \BibitemOpen
  \bibfield  {author} {\bibinfo {author} {\bibfnamefont {C.}~\bibnamefont
  {Becker}}, \bibinfo {author} {\bibfnamefont {S.}~\bibnamefont {Stellmer}},
  \bibinfo {author} {\bibfnamefont {P.}~\bibnamefont {Soltan-Panahi}}, \bibinfo
  {author} {\bibfnamefont {S.}~\bibnamefont {Dörscher}}, \bibinfo {author}
  {\bibfnamefont {M.}~\bibnamefont {Baumert}}, \bibinfo {author} {\bibfnamefont
  {E.-M.}\ \bibnamefont {Richter}}, \bibinfo {author} {\bibfnamefont
  {J.}~\bibnamefont {Kronjäger}}, \bibinfo {author} {\bibfnamefont
  {K.}~\bibnamefont {Bongs}},\ and\ \bibinfo {author} {\bibfnamefont
  {K.}~\bibnamefont {Sengstock}},\ }\href {https://doi.org/10.1038/nphys962}
  {\bibfield  {journal} {\bibinfo  {journal} {Nature Physics}\ }\textbf
  {\bibinfo {volume} {4}},\ \bibinfo {pages} {496} (\bibinfo {year} {2008})},\
  \bibinfo {note} {number: 6 Publisher: Nature Publishing Group}\BibitemShut
  {NoStop}%
\bibitem [{\citenamefont {Chabchoub}\ \emph {et~al.}(2016)\citenamefont
  {Chabchoub}, \citenamefont {Onorato},\ and\ \citenamefont
  {Akhmediev}}]{chabchoub_hydrodynamic_2016}%
  \BibitemOpen
  \bibfield  {author} {\bibinfo {author} {\bibfnamefont {A.}~\bibnamefont
  {Chabchoub}}, \bibinfo {author} {\bibfnamefont {M.}~\bibnamefont {Onorato}},\
  and\ \bibinfo {author} {\bibfnamefont {N.}~\bibnamefont {Akhmediev}},\ }in\
  \href {https://doi.org/10.1007/978-3-319-39214-1_3} {\emph {\bibinfo
  {booktitle} {Rogue and {Shock} {Waves} in {Nonlinear} {Dispersive}
  {Media}}}},\ \bibinfo {series and number} {Lecture {Notes} in {Physics}},\
  \bibinfo {editor} {edited by\ \bibinfo {editor} {\bibfnamefont
  {M.}~\bibnamefont {Onorato}}, \bibinfo {editor} {\bibfnamefont
  {S.}~\bibnamefont {Resitori}},\ and\ \bibinfo {editor} {\bibfnamefont
  {F.}~\bibnamefont {Baronio}}}\ (\bibinfo  {publisher} {Springer International
  Publishing},\ \bibinfo {address} {Cham},\ \bibinfo {year} {2016})\ pp.\
  \bibinfo {pages} {55--87}\BibitemShut {NoStop}%
\bibitem [{\citenamefont {Amo}\ \emph {et~al.}(2011)\citenamefont {Amo},
  \citenamefont {Pigeon}, \citenamefont {Sanvitto}, \citenamefont {Sala},
  \citenamefont {Hivet}, \citenamefont {Carusotto}, \citenamefont {Pisanello},
  \citenamefont {Leménager}, \citenamefont {Houdré}, \citenamefont
  {Giacobino}, \citenamefont {Ciuti},\ and\ \citenamefont
  {Bramati}}]{amo_polariton_2011}%
  \BibitemOpen
  \bibfield  {author} {\bibinfo {author} {\bibfnamefont {A.}~\bibnamefont
  {Amo}}, \bibinfo {author} {\bibfnamefont {S.}~\bibnamefont {Pigeon}},
  \bibinfo {author} {\bibfnamefont {D.}~\bibnamefont {Sanvitto}}, \bibinfo
  {author} {\bibfnamefont {V.~G.}\ \bibnamefont {Sala}}, \bibinfo {author}
  {\bibfnamefont {R.}~\bibnamefont {Hivet}}, \bibinfo {author} {\bibfnamefont
  {I.}~\bibnamefont {Carusotto}}, \bibinfo {author} {\bibfnamefont
  {F.}~\bibnamefont {Pisanello}}, \bibinfo {author} {\bibfnamefont
  {G.}~\bibnamefont {Leménager}}, \bibinfo {author} {\bibfnamefont
  {R.}~\bibnamefont {Houdré}}, \bibinfo {author} {\bibfnamefont
  {E.}~\bibnamefont {Giacobino}}, \bibinfo {author} {\bibfnamefont
  {C.}~\bibnamefont {Ciuti}},\ and\ \bibinfo {author} {\bibfnamefont
  {A.}~\bibnamefont {Bramati}},\ }\href
  {https://doi.org/10.1126/science.1202307} {\bibfield  {journal} {\bibinfo
  {journal} {Science}\ }\textbf {\bibinfo {volume} {332}},\ \bibinfo {pages}
  {1167} (\bibinfo {year} {2011})},\ \bibinfo {note} {publisher: American
  Association for the Advancement of Science Section: Report}\BibitemShut
  {NoStop}%
\bibitem [{\citenamefont {Grelu}\ and\ \citenamefont
  {Akhmediev}(2012)}]{grelu_dissipative_2012}%
  \BibitemOpen
  \bibfield  {author} {\bibinfo {author} {\bibfnamefont {P.}~\bibnamefont
  {Grelu}}\ and\ \bibinfo {author} {\bibfnamefont {N.}~\bibnamefont
  {Akhmediev}},\ }\href {https://doi.org/10.1038/nphoton.2011.345} {\bibfield
  {journal} {\bibinfo  {journal} {Nature Photonics}\ }\textbf {\bibinfo
  {volume} {6}},\ \bibinfo {pages} {84} (\bibinfo {year} {2012})}\BibitemShut
  {NoStop}%
\bibitem [{\citenamefont {Liu}\ \emph {et~al.}(2015)\citenamefont {Liu},
  \citenamefont {Pang}, \citenamefont {Han}, \citenamefont {Tian},
  \citenamefont {Chen}, \citenamefont {Lei}, \citenamefont {Yan},\ and\
  \citenamefont {Wei}}]{liu_generation_2015}%
  \BibitemOpen
  \bibfield  {author} {\bibinfo {author} {\bibfnamefont {W.}~\bibnamefont
  {Liu}}, \bibinfo {author} {\bibfnamefont {L.}~\bibnamefont {Pang}}, \bibinfo
  {author} {\bibfnamefont {H.}~\bibnamefont {Han}}, \bibinfo {author}
  {\bibfnamefont {W.}~\bibnamefont {Tian}}, \bibinfo {author} {\bibfnamefont
  {H.}~\bibnamefont {Chen}}, \bibinfo {author} {\bibfnamefont {M.}~\bibnamefont
  {Lei}}, \bibinfo {author} {\bibfnamefont {P.}~\bibnamefont {Yan}},\ and\
  \bibinfo {author} {\bibfnamefont {Z.}~\bibnamefont {Wei}},\ }\href
  {https://doi.org/10.1364/OE.23.026023} {\bibfield  {journal} {\bibinfo
  {journal} {Optics Express}\ }\textbf {\bibinfo {volume} {23}},\ \bibinfo
  {pages} {26023} (\bibinfo {year} {2015})},\ \bibinfo {note} {publisher:
  Optical Society of America}\BibitemShut {NoStop}%
\bibitem [{\citenamefont {Kippenberg}\ \emph {et~al.}(2018)\citenamefont
  {Kippenberg}, \citenamefont {Gaeta}, \citenamefont {Lipson},\ and\
  \citenamefont {Gorodetsky}}]{kippenberg_dissipative_2018}%
  \BibitemOpen
  \bibfield  {author} {\bibinfo {author} {\bibfnamefont {T.~J.}\ \bibnamefont
  {Kippenberg}}, \bibinfo {author} {\bibfnamefont {A.~L.}\ \bibnamefont
  {Gaeta}}, \bibinfo {author} {\bibfnamefont {M.}~\bibnamefont {Lipson}},\ and\
  \bibinfo {author} {\bibfnamefont {M.~L.}\ \bibnamefont {Gorodetsky}},\ }\href
  {https://doi.org/10.1126/science.aan8083} {\bibfield  {journal} {\bibinfo
  {journal} {Science}\ }\textbf {\bibinfo {volume} {361}},\ \bibinfo {pages}
  {eaan8083} (\bibinfo {year} {2018})}\BibitemShut {NoStop}%
\bibitem [{\citenamefont {Wai}\ \emph {et~al.}(1987)\citenamefont {Wai},
  \citenamefont {Menyuk}, \citenamefont {Chen},\ and\ \citenamefont
  {Lee}}]{wai_soliton_1987}%
  \BibitemOpen
  \bibfield  {author} {\bibinfo {author} {\bibfnamefont {P.~K.~A.}\
  \bibnamefont {Wai}}, \bibinfo {author} {\bibfnamefont {C.~R.}\ \bibnamefont
  {Menyuk}}, \bibinfo {author} {\bibfnamefont {H.~H.}\ \bibnamefont {Chen}},\
  and\ \bibinfo {author} {\bibfnamefont {Y.~C.}\ \bibnamefont {Lee}},\ }\href
  {https://doi.org/10.1364/OL.12.000628} {\bibfield  {journal} {\bibinfo
  {journal} {Optics Letters}\ }\textbf {\bibinfo {volume} {12}},\ \bibinfo
  {pages} {628} (\bibinfo {year} {1987})},\ \bibinfo {note} {publisher: Optical
  Society of America}\BibitemShut {NoStop}%
\bibitem [{\citenamefont {Parra-Rivas}\ \emph {et~al.}(2017)\citenamefont
  {Parra-Rivas}, \citenamefont {Gomila},\ and\ \citenamefont
  {Gelens}}]{parra-rivas_coexistence_2017}%
  \BibitemOpen
  \bibfield  {author} {\bibinfo {author} {\bibfnamefont {P.}~\bibnamefont
  {Parra-Rivas}}, \bibinfo {author} {\bibfnamefont {D.}~\bibnamefont
  {Gomila}},\ and\ \bibinfo {author} {\bibfnamefont {L.}~\bibnamefont
  {Gelens}},\ }\href {https://doi.org/10.1103/PhysRevA.95.053863} {\bibfield
  {journal} {\bibinfo  {journal} {Physical Review A}\ }\textbf {\bibinfo
  {volume} {95}},\ \bibinfo {pages} {053863} (\bibinfo {year}
  {2017})}\BibitemShut {NoStop}%
\bibitem [{\citenamefont {Lobanov}\ \emph {et~al.}(2019)\citenamefont
  {Lobanov}, \citenamefont {Kondratiev}, \citenamefont {Shitikov},
  \citenamefont {Galiev},\ and\ \citenamefont
  {Bilenko}}]{lobanov_generation_2019}%
  \BibitemOpen
  \bibfield  {author} {\bibinfo {author} {\bibfnamefont {V.~E.}\ \bibnamefont
  {Lobanov}}, \bibinfo {author} {\bibfnamefont {N.~M.}\ \bibnamefont
  {Kondratiev}}, \bibinfo {author} {\bibfnamefont {A.~E.}\ \bibnamefont
  {Shitikov}}, \bibinfo {author} {\bibfnamefont {R.~R.}\ \bibnamefont
  {Galiev}},\ and\ \bibinfo {author} {\bibfnamefont {I.~A.}\ \bibnamefont
  {Bilenko}},\ }\href {https://doi.org/10.1103/PhysRevA.100.013807} {\bibfield
  {journal} {\bibinfo  {journal} {Physical Review A}\ }\textbf {\bibinfo
  {volume} {100}},\ \bibinfo {pages} {013807} (\bibinfo {year} {2019})},\
  \bibinfo {note} {publisher: American Physical Society}\BibitemShut {NoStop}%
\bibitem [{\citenamefont {Obrzud}\ \emph {et~al.}(2017)\citenamefont {Obrzud},
  \citenamefont {Lecomte},\ and\ \citenamefont {Herr}}]{obrzud_temporal_2017}%
  \BibitemOpen
  \bibfield  {author} {\bibinfo {author} {\bibfnamefont {E.}~\bibnamefont
  {Obrzud}}, \bibinfo {author} {\bibfnamefont {S.}~\bibnamefont {Lecomte}},\
  and\ \bibinfo {author} {\bibfnamefont {T.}~\bibnamefont {Herr}},\ }\href
  {https://doi.org/10.1038/nphoton.2017.140} {\bibfield  {journal} {\bibinfo
  {journal} {Nature Photonics}\ }\textbf {\bibinfo {volume} {11}},\ \bibinfo
  {pages} {nphoton.2017.140} (\bibinfo {year} {2017})}\BibitemShut {NoStop}%
\bibitem [{\citenamefont {Copie}\ \emph {et~al.}(2016)\citenamefont {Copie},
  \citenamefont {Conforti}, \citenamefont {Kudlinski}, \citenamefont {Mussot},\
  and\ \citenamefont {Trillo}}]{copie_competing_2016}%
  \BibitemOpen
  \bibfield  {author} {\bibinfo {author} {\bibfnamefont {F.}~\bibnamefont
  {Copie}}, \bibinfo {author} {\bibfnamefont {M.}~\bibnamefont {Conforti}},
  \bibinfo {author} {\bibfnamefont {A.}~\bibnamefont {Kudlinski}}, \bibinfo
  {author} {\bibfnamefont {A.}~\bibnamefont {Mussot}},\ and\ \bibinfo {author}
  {\bibfnamefont {S.}~\bibnamefont {Trillo}},\ }\bibfield  {journal} {\bibinfo
  {journal} {Physical Review Letters}\ }\textbf {\bibinfo {volume} {116}},\
  \href {https://doi.org/10.1103/PhysRevLett.116.143901}
  {10.1103/PhysRevLett.116.143901} (\bibinfo {year} {2016})\BibitemShut
  {NoStop}%
\bibitem [{\citenamefont {Herr}\ \emph {et~al.}(2014)\citenamefont {Herr},
  \citenamefont {Brasch}, \citenamefont {Jost}, \citenamefont {Wang},
  \citenamefont {Kondratiev}, \citenamefont {Gorodetsky},\ and\ \citenamefont
  {Kippenberg}}]{herr_temporal_2014}%
  \BibitemOpen
  \bibfield  {author} {\bibinfo {author} {\bibfnamefont {T.}~\bibnamefont
  {Herr}}, \bibinfo {author} {\bibfnamefont {V.}~\bibnamefont {Brasch}},
  \bibinfo {author} {\bibfnamefont {J.~D.}\ \bibnamefont {Jost}}, \bibinfo
  {author} {\bibfnamefont {C.~Y.}\ \bibnamefont {Wang}}, \bibinfo {author}
  {\bibfnamefont {N.~M.}\ \bibnamefont {Kondratiev}}, \bibinfo {author}
  {\bibfnamefont {M.~L.}\ \bibnamefont {Gorodetsky}},\ and\ \bibinfo {author}
  {\bibfnamefont {T.~J.}\ \bibnamefont {Kippenberg}},\ }\href
  {https://doi.org/10.1038/nphoton.2013.343} {\bibfield  {journal} {\bibinfo
  {journal} {Nature Photonics}\ }\textbf {\bibinfo {volume} {8}},\ \bibinfo
  {pages} {145} (\bibinfo {year} {2014})}\BibitemShut {NoStop}%
\bibitem [{\citenamefont {Leo}\ \emph {et~al.}(2010)\citenamefont {Leo},
  \citenamefont {Coen}, \citenamefont {Kockaert}, \citenamefont {Gorza},
  \citenamefont {Emplit},\ and\ \citenamefont
  {Haelterman}}]{leo_temporal_2010}%
  \BibitemOpen
  \bibfield  {author} {\bibinfo {author} {\bibfnamefont {F.}~\bibnamefont
  {Leo}}, \bibinfo {author} {\bibfnamefont {S.}~\bibnamefont {Coen}}, \bibinfo
  {author} {\bibfnamefont {P.}~\bibnamefont {Kockaert}}, \bibinfo {author}
  {\bibfnamefont {S.-P.}\ \bibnamefont {Gorza}}, \bibinfo {author}
  {\bibfnamefont {P.}~\bibnamefont {Emplit}},\ and\ \bibinfo {author}
  {\bibfnamefont {M.}~\bibnamefont {Haelterman}},\ }\href
  {https://doi.org/10.1038/nphoton.2010.120} {\bibfield  {journal} {\bibinfo
  {journal} {Nature Photonics}\ }\textbf {\bibinfo {volume} {4}},\ \bibinfo
  {pages} {471} (\bibinfo {year} {2010})}\BibitemShut {NoStop}%
\bibitem [{\citenamefont {Marin-Palomo}\ \emph {et~al.}(2017)\citenamefont
  {Marin-Palomo}, \citenamefont {Kemal}, \citenamefont {Karpov}, \citenamefont
  {Kordts}, \citenamefont {Pfeifle}, \citenamefont {Pfeiffer}, \citenamefont
  {Trocha}, \citenamefont {Wolf}, \citenamefont {Brasch}, \citenamefont
  {Anderson}, \citenamefont {Rosenberger}, \citenamefont {Vijayan},
  \citenamefont {Freude}, \citenamefont {Kippenberg},\ and\ \citenamefont
  {Koos}}]{marin-palomo_microresonator-based_2017}%
  \BibitemOpen
  \bibfield  {author} {\bibinfo {author} {\bibfnamefont {P.}~\bibnamefont
  {Marin-Palomo}}, \bibinfo {author} {\bibfnamefont {J.~N.}\ \bibnamefont
  {Kemal}}, \bibinfo {author} {\bibfnamefont {M.}~\bibnamefont {Karpov}},
  \bibinfo {author} {\bibfnamefont {A.}~\bibnamefont {Kordts}}, \bibinfo
  {author} {\bibfnamefont {J.}~\bibnamefont {Pfeifle}}, \bibinfo {author}
  {\bibfnamefont {M.~H.~P.}\ \bibnamefont {Pfeiffer}}, \bibinfo {author}
  {\bibfnamefont {P.}~\bibnamefont {Trocha}}, \bibinfo {author} {\bibfnamefont
  {S.}~\bibnamefont {Wolf}}, \bibinfo {author} {\bibfnamefont {V.}~\bibnamefont
  {Brasch}}, \bibinfo {author} {\bibfnamefont {M.~H.}\ \bibnamefont
  {Anderson}}, \bibinfo {author} {\bibfnamefont {R.}~\bibnamefont
  {Rosenberger}}, \bibinfo {author} {\bibfnamefont {K.}~\bibnamefont
  {Vijayan}}, \bibinfo {author} {\bibfnamefont {W.}~\bibnamefont {Freude}},
  \bibinfo {author} {\bibfnamefont {T.~J.}\ \bibnamefont {Kippenberg}},\ and\
  \bibinfo {author} {\bibfnamefont {C.}~\bibnamefont {Koos}},\ }\href
  {https://doi.org/10.1038/nature22387} {\bibfield  {journal} {\bibinfo
  {journal} {Nature}\ }\textbf {\bibinfo {volume} {546}},\ \bibinfo {pages}
  {274} (\bibinfo {year} {2017})}\BibitemShut {NoStop}%
\bibitem [{\citenamefont {Riemensberger}\ \emph {et~al.}(2020)\citenamefont
  {Riemensberger}, \citenamefont {Lukashchuk}, \citenamefont {Karpov},
  \citenamefont {Weng}, \citenamefont {Lucas}, \citenamefont {Liu},\ and\
  \citenamefont {Kippenberg}}]{riemensberger_massively_2020}%
  \BibitemOpen
  \bibfield  {author} {\bibinfo {author} {\bibfnamefont {J.}~\bibnamefont
  {Riemensberger}}, \bibinfo {author} {\bibfnamefont {A.}~\bibnamefont
  {Lukashchuk}}, \bibinfo {author} {\bibfnamefont {M.}~\bibnamefont {Karpov}},
  \bibinfo {author} {\bibfnamefont {W.}~\bibnamefont {Weng}}, \bibinfo {author}
  {\bibfnamefont {E.}~\bibnamefont {Lucas}}, \bibinfo {author} {\bibfnamefont
  {J.}~\bibnamefont {Liu}},\ and\ \bibinfo {author} {\bibfnamefont {T.~J.}\
  \bibnamefont {Kippenberg}},\ }\href
  {https://doi.org/10.1038/s41586-020-2239-3} {\bibfield  {journal} {\bibinfo
  {journal} {Nature}\ }\textbf {\bibinfo {volume} {581}},\ \bibinfo {pages}
  {164} (\bibinfo {year} {2020})},\ \bibinfo {note} {number: 7807 Publisher:
  Nature Publishing Group}\BibitemShut {NoStop}%
\bibitem [{\citenamefont {{Ewelina Obrzud}}\ \emph {et~al.}(2019)\citenamefont
  {{Ewelina Obrzud}}, \citenamefont {{Monica Rainer}}, \citenamefont {{Avet
  Harutyunyan}}, \citenamefont {{Miles H. Anderson}}, \citenamefont {{Junqiu
  Liu}}, \citenamefont {{Michael Geiselmann}}, \citenamefont {{Bruno
  Chazelas}}, \citenamefont {{Stefan Kundermann}}, \citenamefont {{Steve
  Lecomte}}, \citenamefont {{Massimo Cecconi}}, \citenamefont {{Adriano
  Ghedina}}, \citenamefont {{Emilio Molinari}}, \citenamefont {{Francesco
  Pepe}}, \citenamefont {{François Wildi}}, \citenamefont {{François
  Bouchy}}, \citenamefont {{Tobias J. Kippenberg}},\ and\ \citenamefont
  {{Tobias Herr}}}]{ewelina_obrzud_microphotonic_2019}%
  \BibitemOpen
  \bibfield  {author} {\bibinfo {author} {\bibnamefont {{Ewelina Obrzud}}},
  \bibinfo {author} {\bibnamefont {{Monica Rainer}}}, \bibinfo {author}
  {\bibnamefont {{Avet Harutyunyan}}}, \bibinfo {author} {\bibnamefont {{Miles
  H. Anderson}}}, \bibinfo {author} {\bibnamefont {{Junqiu Liu}}}, \bibinfo
  {author} {\bibnamefont {{Michael Geiselmann}}}, \bibinfo {author}
  {\bibnamefont {{Bruno Chazelas}}}, \bibinfo {author} {\bibnamefont {{Stefan
  Kundermann}}}, \bibinfo {author} {\bibnamefont {{Steve Lecomte}}}, \bibinfo
  {author} {\bibnamefont {{Massimo Cecconi}}}, \bibinfo {author} {\bibnamefont
  {{Adriano Ghedina}}}, \bibinfo {author} {\bibnamefont {{Emilio Molinari}}},
  \bibinfo {author} {\bibnamefont {{Francesco Pepe}}}, \bibinfo {author}
  {\bibnamefont {{François Wildi}}}, \bibinfo {author} {\bibnamefont
  {{François Bouchy}}}, \bibinfo {author} {\bibnamefont {{Tobias J.
  Kippenberg}}},\ and\ \bibinfo {author} {\bibnamefont {{Tobias Herr}}},\
  }\href {https://doi.org/10.1038/s41566-018-0309-y} {\bibfield  {journal}
  {\bibinfo  {journal} {Nature Photonics}\ }\textbf {\bibinfo {volume} {13}},\
  \bibinfo {pages} {31} (\bibinfo {year} {2019})}\BibitemShut {NoStop}%
\bibitem [{\citenamefont {Suh}\ \emph {et~al.}(2019)\citenamefont {Suh},
  \citenamefont {Yi}, \citenamefont {Lai}, \citenamefont {Leifer},
  \citenamefont {Grudinin}, \citenamefont {Vasisht}, \citenamefont {Martin},
  \citenamefont {Fitzgerald}, \citenamefont {Doppmann}, \citenamefont {Wang},
  \citenamefont {Mawet}, \citenamefont {Papp}, \citenamefont {Diddams},
  \citenamefont {Beichman},\ and\ \citenamefont {Vahala}}]{suh_searching_2019}%
  \BibitemOpen
  \bibfield  {author} {\bibinfo {author} {\bibfnamefont {M.-G.}\ \bibnamefont
  {Suh}}, \bibinfo {author} {\bibfnamefont {X.}~\bibnamefont {Yi}}, \bibinfo
  {author} {\bibfnamefont {Y.-H.}\ \bibnamefont {Lai}}, \bibinfo {author}
  {\bibfnamefont {S.}~\bibnamefont {Leifer}}, \bibinfo {author} {\bibfnamefont
  {I.~S.}\ \bibnamefont {Grudinin}}, \bibinfo {author} {\bibfnamefont
  {G.}~\bibnamefont {Vasisht}}, \bibinfo {author} {\bibfnamefont {E.~C.}\
  \bibnamefont {Martin}}, \bibinfo {author} {\bibfnamefont {M.~P.}\
  \bibnamefont {Fitzgerald}}, \bibinfo {author} {\bibfnamefont
  {G.}~\bibnamefont {Doppmann}}, \bibinfo {author} {\bibfnamefont
  {J.}~\bibnamefont {Wang}}, \bibinfo {author} {\bibfnamefont {D.}~\bibnamefont
  {Mawet}}, \bibinfo {author} {\bibfnamefont {S.~B.}\ \bibnamefont {Papp}},
  \bibinfo {author} {\bibfnamefont {S.~A.}\ \bibnamefont {Diddams}}, \bibinfo
  {author} {\bibfnamefont {C.}~\bibnamefont {Beichman}},\ and\ \bibinfo
  {author} {\bibfnamefont {K.}~\bibnamefont {Vahala}},\ }\href
  {https://doi.org/10.1038/s41566-018-0312-3} {\bibfield  {journal} {\bibinfo
  {journal} {Nature Photonics}\ }\textbf {\bibinfo {volume} {13}},\ \bibinfo
  {pages} {25} (\bibinfo {year} {2019})}\BibitemShut {NoStop}%
\bibitem [{\citenamefont {Suh}\ \emph {et~al.}(2016)\citenamefont {Suh},
  \citenamefont {Yang}, \citenamefont {Yang}, \citenamefont {Yi},\ and\
  \citenamefont {Vahala}}]{suh_microresonator_2016}%
  \BibitemOpen
  \bibfield  {author} {\bibinfo {author} {\bibfnamefont {M.-G.}\ \bibnamefont
  {Suh}}, \bibinfo {author} {\bibfnamefont {Q.-F.}\ \bibnamefont {Yang}},
  \bibinfo {author} {\bibfnamefont {K.~Y.}\ \bibnamefont {Yang}}, \bibinfo
  {author} {\bibfnamefont {X.}~\bibnamefont {Yi}},\ and\ \bibinfo {author}
  {\bibfnamefont {K.~J.}\ \bibnamefont {Vahala}},\ }\href
  {https://doi.org/10.1126/science.aah6516} {\bibfield  {journal} {\bibinfo
  {journal} {Science}\ }\textbf {\bibinfo {volume} {354}},\ \bibinfo {pages}
  {600} (\bibinfo {year} {2016})}\BibitemShut {NoStop}%
\bibitem [{\citenamefont {Spencer}\ \emph {et~al.}(2018)\citenamefont
  {Spencer}, \citenamefont {Drake}, \citenamefont {Briles}, \citenamefont
  {Stone}, \citenamefont {Sinclair}, \citenamefont {Fredrick}, \citenamefont
  {Li}, \citenamefont {Westly}, \citenamefont {Ilic}, \citenamefont
  {Bluestone}, \citenamefont {Volet}, \citenamefont {Komljenovic},
  \citenamefont {Chang}, \citenamefont {Lee}, \citenamefont {Oh}, \citenamefont
  {Suh}, \citenamefont {Yang}, \citenamefont {Pfeiffer}, \citenamefont
  {Kippenberg}, \citenamefont {Norberg}, \citenamefont {Theogarajan},
  \citenamefont {Vahala}, \citenamefont {Newbury}, \citenamefont {Srinivasan},
  \citenamefont {Bowers}, \citenamefont {Diddams},\ and\ \citenamefont
  {Papp}}]{spencer_optical-frequency_2018}%
  \BibitemOpen
  \bibfield  {author} {\bibinfo {author} {\bibfnamefont {D.~T.}\ \bibnamefont
  {Spencer}}, \bibinfo {author} {\bibfnamefont {T.}~\bibnamefont {Drake}},
  \bibinfo {author} {\bibfnamefont {T.~C.}\ \bibnamefont {Briles}}, \bibinfo
  {author} {\bibfnamefont {J.}~\bibnamefont {Stone}}, \bibinfo {author}
  {\bibfnamefont {L.~C.}\ \bibnamefont {Sinclair}}, \bibinfo {author}
  {\bibfnamefont {C.}~\bibnamefont {Fredrick}}, \bibinfo {author}
  {\bibfnamefont {Q.}~\bibnamefont {Li}}, \bibinfo {author} {\bibfnamefont
  {D.}~\bibnamefont {Westly}}, \bibinfo {author} {\bibfnamefont {B.~R.}\
  \bibnamefont {Ilic}}, \bibinfo {author} {\bibfnamefont {A.}~\bibnamefont
  {Bluestone}}, \bibinfo {author} {\bibfnamefont {N.}~\bibnamefont {Volet}},
  \bibinfo {author} {\bibfnamefont {T.}~\bibnamefont {Komljenovic}}, \bibinfo
  {author} {\bibfnamefont {L.}~\bibnamefont {Chang}}, \bibinfo {author}
  {\bibfnamefont {S.~H.}\ \bibnamefont {Lee}}, \bibinfo {author} {\bibfnamefont
  {D.~Y.}\ \bibnamefont {Oh}}, \bibinfo {author} {\bibfnamefont {M.-G.}\
  \bibnamefont {Suh}}, \bibinfo {author} {\bibfnamefont {K.~Y.}\ \bibnamefont
  {Yang}}, \bibinfo {author} {\bibfnamefont {M.~H.~P.}\ \bibnamefont
  {Pfeiffer}}, \bibinfo {author} {\bibfnamefont {T.~J.}\ \bibnamefont
  {Kippenberg}}, \bibinfo {author} {\bibfnamefont {E.}~\bibnamefont {Norberg}},
  \bibinfo {author} {\bibfnamefont {L.}~\bibnamefont {Theogarajan}}, \bibinfo
  {author} {\bibfnamefont {K.}~\bibnamefont {Vahala}}, \bibinfo {author}
  {\bibfnamefont {N.~R.}\ \bibnamefont {Newbury}}, \bibinfo {author}
  {\bibfnamefont {K.}~\bibnamefont {Srinivasan}}, \bibinfo {author}
  {\bibfnamefont {J.~E.}\ \bibnamefont {Bowers}}, \bibinfo {author}
  {\bibfnamefont {S.~A.}\ \bibnamefont {Diddams}},\ and\ \bibinfo {author}
  {\bibfnamefont {S.~B.}\ \bibnamefont {Papp}},\ }\href
  {https://doi.org/10.1038/s41586-018-0065-7} {\bibfield  {journal} {\bibinfo
  {journal} {Nature}\ ,\ \bibinfo {pages} {1}} (\bibinfo {year}
  {2018})}\BibitemShut {NoStop}%
\bibitem [{\citenamefont {Newman}\ \emph {et~al.}(2019)\citenamefont {Newman},
  \citenamefont {Maurice}, \citenamefont {Drake}, \citenamefont {Stone},
  \citenamefont {Briles}, \citenamefont {Spencer}, \citenamefont {Fredrick},
  \citenamefont {Li}, \citenamefont {Westly}, \citenamefont {Ilic},
  \citenamefont {Shen}, \citenamefont {Suh}, \citenamefont {Yang},
  \citenamefont {Johnson}, \citenamefont {Johnson}, \citenamefont {Hollberg},
  \citenamefont {Vahala}, \citenamefont {Srinivasan}, \citenamefont {Diddams},
  \citenamefont {Kitching}, \citenamefont {Papp},\ and\ \citenamefont
  {Hummon}}]{newman_architecture_2019}%
  \BibitemOpen
  \bibfield  {author} {\bibinfo {author} {\bibfnamefont {Z.~L.}\ \bibnamefont
  {Newman}}, \bibinfo {author} {\bibfnamefont {V.}~\bibnamefont {Maurice}},
  \bibinfo {author} {\bibfnamefont {T.}~\bibnamefont {Drake}}, \bibinfo
  {author} {\bibfnamefont {J.~R.}\ \bibnamefont {Stone}}, \bibinfo {author}
  {\bibfnamefont {T.~C.}\ \bibnamefont {Briles}}, \bibinfo {author}
  {\bibfnamefont {D.~T.}\ \bibnamefont {Spencer}}, \bibinfo {author}
  {\bibfnamefont {C.}~\bibnamefont {Fredrick}}, \bibinfo {author}
  {\bibfnamefont {Q.}~\bibnamefont {Li}}, \bibinfo {author} {\bibfnamefont
  {D.}~\bibnamefont {Westly}}, \bibinfo {author} {\bibfnamefont {B.~R.}\
  \bibnamefont {Ilic}}, \bibinfo {author} {\bibfnamefont {B.}~\bibnamefont
  {Shen}}, \bibinfo {author} {\bibfnamefont {M.-G.}\ \bibnamefont {Suh}},
  \bibinfo {author} {\bibfnamefont {K.~Y.}\ \bibnamefont {Yang}}, \bibinfo
  {author} {\bibfnamefont {C.}~\bibnamefont {Johnson}}, \bibinfo {author}
  {\bibfnamefont {D.~M.~S.}\ \bibnamefont {Johnson}}, \bibinfo {author}
  {\bibfnamefont {L.}~\bibnamefont {Hollberg}}, \bibinfo {author}
  {\bibfnamefont {K.~J.}\ \bibnamefont {Vahala}}, \bibinfo {author}
  {\bibfnamefont {K.}~\bibnamefont {Srinivasan}}, \bibinfo {author}
  {\bibfnamefont {S.~A.}\ \bibnamefont {Diddams}}, \bibinfo {author}
  {\bibfnamefont {J.}~\bibnamefont {Kitching}}, \bibinfo {author}
  {\bibfnamefont {S.~B.}\ \bibnamefont {Papp}},\ and\ \bibinfo {author}
  {\bibfnamefont {M.~T.}\ \bibnamefont {Hummon}},\ }\href
  {https://doi.org/10.1364/OPTICA.6.000680} {\bibfield  {journal} {\bibinfo
  {journal} {Optica}\ }\textbf {\bibinfo {volume} {6}},\ \bibinfo {pages} {680}
  (\bibinfo {year} {2019})}\BibitemShut {NoStop}%
\bibitem [{\citenamefont {Lugiato}\ and\ \citenamefont
  {Lefever}(1987)}]{lugiato_spatial_1987}%
  \BibitemOpen
  \bibfield  {author} {\bibinfo {author} {\bibfnamefont {L.~A.}\ \bibnamefont
  {Lugiato}}\ and\ \bibinfo {author} {\bibfnamefont {R.}~\bibnamefont
  {Lefever}},\ }\href {https://doi.org/10.1103/PhysRevLett.58.2209} {\bibfield
  {journal} {\bibinfo  {journal} {Physical Review Letters}\ }\textbf {\bibinfo
  {volume} {58}},\ \bibinfo {pages} {2209} (\bibinfo {year}
  {1987})}\BibitemShut {NoStop}%
\bibitem [{\citenamefont {Ackemann}\ and\ \citenamefont
  {Firth}(2005)}]{ackemann_dissipative_2005}%
  \BibitemOpen
  \bibfield  {author} {\bibinfo {author} {\bibfnamefont {T.}~\bibnamefont
  {Ackemann}}\ and\ \bibinfo {author} {\bibfnamefont {W.}~\bibnamefont
  {Firth}},\ }in\ \href {https://doi.org/10.1007/10928028_4} {\emph {\bibinfo
  {booktitle} {Dissipative {Solitons}}}},\ \bibinfo {editor} {edited by\
  \bibinfo {editor} {\bibfnamefont {N.}~\bibnamefont {Akhmediev}}\ and\
  \bibinfo {editor} {\bibfnamefont {A.}~\bibnamefont {Ankiewicz}}}\ (\bibinfo
  {publisher} {Springer Berlin Heidelberg},\ \bibinfo {address} {Berlin,
  Heidelberg},\ \bibinfo {year} {2005})\ pp.\ \bibinfo {pages}
  {55--100}\BibitemShut {NoStop}%
\bibitem [{\citenamefont {Nozaki}\ and\ \citenamefont
  {Bekki}(1985)}]{nozaki_chaotic_1985}%
  \BibitemOpen
  \bibfield  {author} {\bibinfo {author} {\bibfnamefont {K.}~\bibnamefont
  {Nozaki}}\ and\ \bibinfo {author} {\bibfnamefont {N.}~\bibnamefont {Bekki}},\
  }\href {https://doi.org/10.1143/JPSJ.54.2363} {\bibfield  {journal} {\bibinfo
   {journal} {Journal of the Physical Society of Japan}\ }\textbf {\bibinfo
  {volume} {54}},\ \bibinfo {pages} {2363} (\bibinfo {year} {1985})},\ \bibinfo
  {note} {publisher: The Physical Society of Japan}\BibitemShut {NoStop}%
\bibitem [{\citenamefont {Gonzalez-Perez}\ \emph {et~al.}(2016)\citenamefont
  {Gonzalez-Perez}, \citenamefont {Mosgaard}, \citenamefont {Budvytyte},
  \citenamefont {Villagran-Vargas}, \citenamefont {Jackson},\ and\
  \citenamefont {Heimburg}}]{gonzalez-perez_solitary_2016}%
  \BibitemOpen
  \bibfield  {author} {\bibinfo {author} {\bibfnamefont {A.}~\bibnamefont
  {Gonzalez-Perez}}, \bibinfo {author} {\bibfnamefont {L.~D.}\ \bibnamefont
  {Mosgaard}}, \bibinfo {author} {\bibfnamefont {R.}~\bibnamefont {Budvytyte}},
  \bibinfo {author} {\bibfnamefont {E.}~\bibnamefont {Villagran-Vargas}},
  \bibinfo {author} {\bibfnamefont {A.~D.}\ \bibnamefont {Jackson}},\ and\
  \bibinfo {author} {\bibfnamefont {T.}~\bibnamefont {Heimburg}},\ }\href
  {https://doi.org/10.1016/j.bpc.2016.06.005} {\bibfield  {journal} {\bibinfo
  {journal} {Biophysical Chemistry}\ }\textbf {\bibinfo {volume} {216}},\
  \bibinfo {pages} {51} (\bibinfo {year} {2016})}\BibitemShut {NoStop}%
\bibitem [{\citenamefont {Liehr}(2013)}]{liehr2013dissipative}%
  \BibitemOpen
  \bibfield  {author} {\bibinfo {author} {\bibfnamefont {A.}~\bibnamefont
  {Liehr}},\ }\href@noop {} {\emph {\bibinfo {title} {Dissipative solitons in
  reaction diffusion systems}}},\ Vol.~\bibinfo {volume} {70}\ (\bibinfo
  {publisher} {Springer},\ \bibinfo {year} {2013})\BibitemShut {NoStop}%
\bibitem [{\citenamefont {Liang}\ \emph {et~al.}(2014)\citenamefont {Liang},
  \citenamefont {Savchenkov}, \citenamefont {Ilchenko}, \citenamefont
  {Eliyahu}, \citenamefont {Seidel}, \citenamefont {Matsko},\ and\
  \citenamefont {Maleki}}]{liang_generation_2014}%
  \BibitemOpen
  \bibfield  {author} {\bibinfo {author} {\bibfnamefont {W.}~\bibnamefont
  {Liang}}, \bibinfo {author} {\bibfnamefont {A.~A.}\ \bibnamefont
  {Savchenkov}}, \bibinfo {author} {\bibfnamefont {V.~S.}\ \bibnamefont
  {Ilchenko}}, \bibinfo {author} {\bibfnamefont {D.}~\bibnamefont {Eliyahu}},
  \bibinfo {author} {\bibfnamefont {D.}~\bibnamefont {Seidel}}, \bibinfo
  {author} {\bibfnamefont {A.~B.}\ \bibnamefont {Matsko}},\ and\ \bibinfo
  {author} {\bibfnamefont {L.}~\bibnamefont {Maleki}},\ }\href
  {https://doi.org/10.1364/OL.39.002920} {\bibfield  {journal} {\bibinfo
  {journal} {Optics Letters}\ }\textbf {\bibinfo {volume} {39}},\ \bibinfo
  {pages} {2920} (\bibinfo {year} {2014})}\BibitemShut {NoStop}%
\bibitem [{\citenamefont {Xue}\ \emph {et~al.}(2015)\citenamefont {Xue},
  \citenamefont {Xuan}, \citenamefont {Liu}, \citenamefont {Wang},
  \citenamefont {Chen}, \citenamefont {Wang}, \citenamefont {Leaird},
  \citenamefont {Qi},\ and\ \citenamefont {Weiner}}]{xue_mode-locked_2015}%
  \BibitemOpen
  \bibfield  {author} {\bibinfo {author} {\bibfnamefont {X.}~\bibnamefont
  {Xue}}, \bibinfo {author} {\bibfnamefont {Y.}~\bibnamefont {Xuan}}, \bibinfo
  {author} {\bibfnamefont {Y.}~\bibnamefont {Liu}}, \bibinfo {author}
  {\bibfnamefont {P.-H.}\ \bibnamefont {Wang}}, \bibinfo {author}
  {\bibfnamefont {S.}~\bibnamefont {Chen}}, \bibinfo {author} {\bibfnamefont
  {J.}~\bibnamefont {Wang}}, \bibinfo {author} {\bibfnamefont {D.~E.}\
  \bibnamefont {Leaird}}, \bibinfo {author} {\bibfnamefont {M.}~\bibnamefont
  {Qi}},\ and\ \bibinfo {author} {\bibfnamefont {A.~M.}\ \bibnamefont
  {Weiner}},\ }\href {https://doi.org/10.1038/nphoton.2015.137} {\bibfield
  {journal} {\bibinfo  {journal} {Nature Photonics}\ }\textbf {\bibinfo
  {volume} {9}},\ \bibinfo {pages} {594} (\bibinfo {year} {2015})}\BibitemShut
  {NoStop}%
\bibitem [{\citenamefont {Huang}\ \emph {et~al.}(2015)\citenamefont {Huang},
  \citenamefont {Zhou}, \citenamefont {Yang}, \citenamefont {McMillan},
  \citenamefont {Matsko}, \citenamefont {Yu}, \citenamefont {Kwong},
  \citenamefont {Maleki},\ and\ \citenamefont {Wong}}]{huang_mode-locked_2015}%
  \BibitemOpen
  \bibfield  {author} {\bibinfo {author} {\bibfnamefont {S.-W.}\ \bibnamefont
  {Huang}}, \bibinfo {author} {\bibfnamefont {H.}~\bibnamefont {Zhou}},
  \bibinfo {author} {\bibfnamefont {J.}~\bibnamefont {Yang}}, \bibinfo {author}
  {\bibfnamefont {J.}~\bibnamefont {McMillan}}, \bibinfo {author}
  {\bibfnamefont {A.}~\bibnamefont {Matsko}}, \bibinfo {author} {\bibfnamefont
  {M.}~\bibnamefont {Yu}}, \bibinfo {author} {\bibfnamefont {D.-L.}\
  \bibnamefont {Kwong}}, \bibinfo {author} {\bibfnamefont {L.}~\bibnamefont
  {Maleki}},\ and\ \bibinfo {author} {\bibfnamefont {C.}~\bibnamefont {Wong}},\
  }\bibfield  {journal} {\bibinfo  {journal} {Physical Review Letters}\
  }\textbf {\bibinfo {volume} {114}},\ \href
  {https://doi.org/10.1103/PhysRevLett.114.053901}
  {10.1103/PhysRevLett.114.053901} (\bibinfo {year} {2015})\BibitemShut
  {NoStop}%
\bibitem [{\citenamefont {Lobanov}\ \emph {et~al.}(2015)\citenamefont
  {Lobanov}, \citenamefont {Lihachev}, \citenamefont {Kippenberg},\ and\
  \citenamefont {Gorodetsky}}]{lobanov_frequency_2015}%
  \BibitemOpen
  \bibfield  {author} {\bibinfo {author} {\bibfnamefont {V.~E.}\ \bibnamefont
  {Lobanov}}, \bibinfo {author} {\bibfnamefont {G.}~\bibnamefont {Lihachev}},
  \bibinfo {author} {\bibfnamefont {T.~J.}\ \bibnamefont {Kippenberg}},\ and\
  \bibinfo {author} {\bibfnamefont {M.~L.}\ \bibnamefont {Gorodetsky}},\ }\href
  {https://doi.org/10.1364/OE.23.007713} {\bibfield  {journal} {\bibinfo
  {journal} {Optics Express}\ }\textbf {\bibinfo {volume} {23}},\ \bibinfo
  {pages} {7713} (\bibinfo {year} {2015})}\BibitemShut {NoStop}%
\bibitem [{\citenamefont {Parra-Rivas}\ \emph {et~al.}(2016)\citenamefont
  {Parra-Rivas}, \citenamefont {Gomila}, \citenamefont {Knobloch},
  \citenamefont {Coen},\ and\ \citenamefont
  {Gelens}}]{parra-rivas_origin_2016}%
  \BibitemOpen
  \bibfield  {author} {\bibinfo {author} {\bibfnamefont {P.}~\bibnamefont
  {Parra-Rivas}}, \bibinfo {author} {\bibfnamefont {D.}~\bibnamefont {Gomila}},
  \bibinfo {author} {\bibfnamefont {E.}~\bibnamefont {Knobloch}}, \bibinfo
  {author} {\bibfnamefont {S.}~\bibnamefont {Coen}},\ and\ \bibinfo {author}
  {\bibfnamefont {L.}~\bibnamefont {Gelens}},\ }\href
  {https://doi.org/10.1364/OL.41.002402} {\bibfield  {journal} {\bibinfo
  {journal} {Optics Letters}\ }\textbf {\bibinfo {volume} {41}},\ \bibinfo
  {pages} {2402} (\bibinfo {year} {2016})}\BibitemShut {NoStop}%
\bibitem [{\citenamefont {Xue}\ \emph {et~al.}(2017)\citenamefont {Xue},
  \citenamefont {Wang}, \citenamefont {Xuan}, \citenamefont {Qi},\ and\
  \citenamefont {Weiner}}]{xue_microresonator_2017}%
  \BibitemOpen
  \bibfield  {author} {\bibinfo {author} {\bibfnamefont {X.}~\bibnamefont
  {Xue}}, \bibinfo {author} {\bibfnamefont {P.-H.}\ \bibnamefont {Wang}},
  \bibinfo {author} {\bibfnamefont {Y.}~\bibnamefont {Xuan}}, \bibinfo {author}
  {\bibfnamefont {M.}~\bibnamefont {Qi}},\ and\ \bibinfo {author}
  {\bibfnamefont {A.~M.}\ \bibnamefont {Weiner}},\ }\href
  {https://doi.org/10.1002/lpor.201600276} {\bibfield  {journal} {\bibinfo
  {journal} {Laser \& Photonics Reviews}\ }\textbf {\bibinfo {volume} {11}},\
  \bibinfo {pages} {1600276} (\bibinfo {year} {2017})},\ \bibinfo {note}
  \BibitemShut
  {NoStop}%
\bibitem [{\citenamefont {F\"{u}l\"{o}p}\ \emph {et~al.}(2018)\citenamefont
  {F\"{u}l\"{o}p}, \citenamefont {Mazur}, \citenamefont {Lorences-Riesgo},
  \citenamefont {Helgason}, \citenamefont {Wang}, \citenamefont {Xuan},
  \citenamefont {Leaird}, \citenamefont {Qi}, \citenamefont {Andrekson},
  \citenamefont {Weiner},\ and\ \citenamefont
  {Torres-Company}}]{fulop_high-order_2018}%
  \BibitemOpen
  \bibfield  {author} {\bibinfo {author} {\bibfnamefont {A.}~\bibnamefont
  {F\"{u}l\"{o}p}}, \bibinfo {author} {\bibfnamefont {M.}~\bibnamefont
  {Mazur}}, \bibinfo {author} {\bibfnamefont {A.}~\bibnamefont
  {Lorences-Riesgo}}, \bibinfo {author} {\bibfnamefont {�.~B.}\ \bibnamefont
  {Helgason}}, \bibinfo {author} {\bibfnamefont {P.-H.}\ \bibnamefont {Wang}},
  \bibinfo {author} {\bibfnamefont {Y.}~\bibnamefont {Xuan}}, \bibinfo {author}
  {\bibfnamefont {D.~E.}\ \bibnamefont {Leaird}}, \bibinfo {author}
  {\bibfnamefont {M.}~\bibnamefont {Qi}}, \bibinfo {author} {\bibfnamefont
  {P.~A.}\ \bibnamefont {Andrekson}}, \bibinfo {author} {\bibfnamefont {A.~M.}\
  \bibnamefont {Weiner}},\ and\ \bibinfo {author} {\bibfnamefont
  {V.}~\bibnamefont {Torres-Company}},\ }\href
  {https://doi.org/10.1038/s41467-018-04046-6} {\bibfield  {journal} {\bibinfo
  {journal} {Nature Communications}\ }\textbf {\bibinfo {volume} {9}},\
  \bibinfo {pages} {1598} (\bibinfo {year} {2018})}\BibitemShut {NoStop}%
\bibitem [{\citenamefont {Rozanov}\ \emph {et~al.}(1982)\citenamefont
  {Rozanov}, \citenamefont {Semenov},\ and\ \citenamefont
  {Khodova}}]{rozanov_transverse_1982}%
  \BibitemOpen
  \bibfield  {author} {\bibinfo {author} {\bibfnamefont {N.~N.}\ \bibnamefont
  {Rozanov}}, \bibinfo {author} {\bibfnamefont {V.~E.}\ \bibnamefont
  {Semenov}},\ and\ \bibinfo {author} {\bibfnamefont {G.~V.}\ \bibnamefont
  {Khodova}},\ }\href {https://doi.org/10.1070/QE1982v012n02ABEH005474}
  {\bibfield  {journal} {\bibinfo  {journal} {Soviet Journal of Quantum
  Electronics}\ }\textbf {\bibinfo {volume} {12}},\ \bibinfo {pages} {193}
  (\bibinfo {year} {1982})},\ \bibinfo {note} {publisher: IOP
  Publishing}\BibitemShut {NoStop}%
\bibitem [{\citenamefont {Trillo}\ \emph {et~al.}(1997)\citenamefont {Trillo},
  \citenamefont {Haelterman},\ and\ \citenamefont
  {Sheppard}}]{trillo_stable_1997}%
  \BibitemOpen
  \bibfield  {author} {\bibinfo {author} {\bibfnamefont {S.}~\bibnamefont
  {Trillo}}, \bibinfo {author} {\bibfnamefont {M.}~\bibnamefont {Haelterman}},\
  and\ \bibinfo {author} {\bibfnamefont {A.}~\bibnamefont {Sheppard}},\ }\href
  {https://doi.org/10.1364/OL.22.000970} {\bibfield  {journal} {\bibinfo
  {journal} {Optics Letters}\ }\textbf {\bibinfo {volume} {22}},\ \bibinfo
  {pages} {970} (\bibinfo {year} {1997})},\ \bibinfo {note} {publisher: Optical
  Society of America}\BibitemShut {NoStop}%
\bibitem [{\citenamefont {Ganne}\ \emph {et~al.}(2001)\citenamefont {Ganne},
  \citenamefont {Slekys}, \citenamefont {Sagnes},\ and\ \citenamefont
  {Kuszelewicz}}]{ganne_optical_2001}%
  \BibitemOpen
  \bibfield  {author} {\bibinfo {author} {\bibfnamefont {I.}~\bibnamefont
  {Ganne}}, \bibinfo {author} {\bibfnamefont {G.}~\bibnamefont {Slekys}},
  \bibinfo {author} {\bibfnamefont {I.}~\bibnamefont {Sagnes}},\ and\ \bibinfo
  {author} {\bibfnamefont {R.}~\bibnamefont {Kuszelewicz}},\ }\href
  {https://doi.org/10.1103/PhysRevB.63.075318} {\bibfield  {journal} {\bibinfo
  {journal} {Physical Review B}\ }\textbf {\bibinfo {volume} {63}},\ \bibinfo
  {pages} {075318} (\bibinfo {year} {2001})},\ \bibinfo {note} {publisher:
  American Physical Society}\BibitemShut {NoStop}%
\bibitem [{\citenamefont {Malomed}(1994)}]{malomed_optical_1994}%
  \BibitemOpen
  \bibfield  {author} {\bibinfo {author} {\bibfnamefont {B.~A.}\ \bibnamefont
  {Malomed}},\ }\href {https://doi.org/10.1103/PhysRevE.50.1565} {\bibfield
  {journal} {\bibinfo  {journal} {Physical Review E}\ }\textbf {\bibinfo
  {volume} {50}},\ \bibinfo {pages} {1565} (\bibinfo {year} {1994})},\ \bibinfo
  {note} {publisher: American Physical Society}\BibitemShut {NoStop}%
\bibitem [{\citenamefont {Garbin}\ \emph {et~al.}(2020)\citenamefont {Garbin},
  \citenamefont {Fatome}, \citenamefont {Oppo}, \citenamefont {Erkintalo},
  \citenamefont {Murdoch},\ and\ \citenamefont
  {Coen}}]{garbin_dissipative_2020}%
  \BibitemOpen
  \bibfield  {author} {\bibinfo {author} {\bibfnamefont {B.}~\bibnamefont
  {Garbin}}, \bibinfo {author} {\bibfnamefont {J.}~\bibnamefont {Fatome}},
  \bibinfo {author} {\bibfnamefont {G.-L.}\ \bibnamefont {Oppo}}, \bibinfo
  {author} {\bibfnamefont {M.}~\bibnamefont {Erkintalo}}, \bibinfo {author}
  {\bibfnamefont {S.~G.}\ \bibnamefont {Murdoch}},\ and\ \bibinfo {author}
  {\bibfnamefont {S.}~\bibnamefont {Coen}},\ }\href
  {http://arxiv.org/abs/2005.09597} {\bibfield  {journal} {\bibinfo  {journal}
  {arXiv:2005.09597 [nlin, physics:physics]}\ } (\bibinfo {year} {2020})},\
  \bibinfo {note} {arXiv: 2005.09597}\BibitemShut {NoStop}%
\bibitem [{\citenamefont {Pomeau}(1986)}]{pomeau_front_1986}%
  \BibitemOpen
  \bibfield  {author} {\bibinfo {author} {\bibfnamefont {Y.}~\bibnamefont
  {Pomeau}},\ }\href {https://doi.org/10.1016/0167-2789(86)90104-1} {\bibfield
  {journal} {\bibinfo  {journal} {Physica D: Nonlinear Phenomena}\ }\textbf
  {\bibinfo {volume} {23}},\ \bibinfo {pages} {3} (\bibinfo {year}
  {1986})}\BibitemShut {NoStop}%
\bibitem [{\citenamefont {Jang}\ \emph {et~al.}(2014)\citenamefont {Jang},
  \citenamefont {Erkintalo}, \citenamefont {Murdoch},\ and\ \citenamefont
  {Coen}}]{jang_observation_2014}%
  \BibitemOpen
  \bibfield  {author} {\bibinfo {author} {\bibfnamefont {J.~K.}\ \bibnamefont
  {Jang}}, \bibinfo {author} {\bibfnamefont {M.}~\bibnamefont {Erkintalo}},
  \bibinfo {author} {\bibfnamefont {S.~G.}\ \bibnamefont {Murdoch}},\ and\
  \bibinfo {author} {\bibfnamefont {S.}~\bibnamefont {Coen}},\ }\href
  {https://doi.org/10.1364/OL.39.005503} {\bibfield  {journal} {\bibinfo
  {journal} {Optics Letters}\ }\textbf {\bibinfo {volume} {39}},\ \bibinfo
  {pages} {5503} (\bibinfo {year} {2014})}\BibitemShut {NoStop}%
\bibitem [{\citenamefont {Brasch}\ \emph {et~al.}(2016)\citenamefont {Brasch},
  \citenamefont {Geiselmann}, \citenamefont {Herr}, \citenamefont {Lihachev},
  \citenamefont {Pfeiffer}, \citenamefont {Gorodetsky},\ and\ \citenamefont
  {Kippenberg}}]{brasch_photonic_2016}%
  \BibitemOpen
  \bibfield  {author} {\bibinfo {author} {\bibfnamefont {V.}~\bibnamefont
  {Brasch}}, \bibinfo {author} {\bibfnamefont {M.}~\bibnamefont {Geiselmann}},
  \bibinfo {author} {\bibfnamefont {T.}~\bibnamefont {Herr}}, \bibinfo {author}
  {\bibfnamefont {G.}~\bibnamefont {Lihachev}}, \bibinfo {author}
  {\bibfnamefont {M.~H.~P.}\ \bibnamefont {Pfeiffer}}, \bibinfo {author}
  {\bibfnamefont {M.~L.}\ \bibnamefont {Gorodetsky}},\ and\ \bibinfo {author}
  {\bibfnamefont {T.~J.}\ \bibnamefont {Kippenberg}},\ }\href
  {https://doi.org/10.1126/science.aad4811} {\bibfield  {journal} {\bibinfo
  {journal} {Science}\ }\textbf {\bibinfo {volume} {351}},\ \bibinfo {pages}
  {357} (\bibinfo {year} {2016})}\BibitemShut {NoStop}%
\bibitem [{\citenamefont {Milián}\ and\ \citenamefont
  {Skryabin}(2014)}]{milian_soliton_2014}%
  \BibitemOpen
  \bibfield  {author} {\bibinfo {author} {\bibfnamefont {C.}~\bibnamefont
  {Milián}}\ and\ \bibinfo {author} {\bibfnamefont {D.~V.}\ \bibnamefont
  {Skryabin}},\ }\href {https://doi.org/10.1364/OE.22.003732} {\bibfield
  {journal} {\bibinfo  {journal} {Optics Express}\ }\textbf {\bibinfo {volume}
  {22}},\ \bibinfo {pages} {3732} (\bibinfo {year} {2014})}\BibitemShut
  {NoStop}%
\bibitem [{\citenamefont {Bao}\ \emph {et~al.}(2017)\citenamefont {Bao},
  \citenamefont {Taheri}, \citenamefont {Zhang}, \citenamefont {Matsko},
  \citenamefont {Yan}, \citenamefont {Liao}, \citenamefont {Maleki},\ and\
  \citenamefont {Willner}}]{bao_high-order_2017}%
  \BibitemOpen
  \bibfield  {author} {\bibinfo {author} {\bibfnamefont {C.}~\bibnamefont
  {Bao}}, \bibinfo {author} {\bibfnamefont {H.}~\bibnamefont {Taheri}},
  \bibinfo {author} {\bibfnamefont {L.}~\bibnamefont {Zhang}}, \bibinfo
  {author} {\bibfnamefont {A.}~\bibnamefont {Matsko}}, \bibinfo {author}
  {\bibfnamefont {Y.}~\bibnamefont {Yan}}, \bibinfo {author} {\bibfnamefont
  {P.}~\bibnamefont {Liao}}, \bibinfo {author} {\bibfnamefont {L.}~\bibnamefont
  {Maleki}},\ and\ \bibinfo {author} {\bibfnamefont {A.~E.}\ \bibnamefont
  {Willner}},\ }\href {https://doi.org/10.1364/JOSAB.34.000715} {\bibfield
  {journal} {\bibinfo  {journal} {JOSA B}\ }\textbf {\bibinfo {volume} {34}},\
  \bibinfo {pages} {715} (\bibinfo {year} {2017})}\BibitemShut {NoStop}%
\bibitem [{\citenamefont {Coen}\ \emph {et~al.}(1999)\citenamefont {Coen},
  \citenamefont {Tlidi}, \citenamefont {Emplit},\ and\ \citenamefont
  {Haelterman}}]{coen_convection_1999}%
  \BibitemOpen
  \bibfield  {author} {\bibinfo {author} {\bibfnamefont {S.}~\bibnamefont
  {Coen}}, \bibinfo {author} {\bibfnamefont {M.}~\bibnamefont {Tlidi}},
  \bibinfo {author} {\bibfnamefont {P.}~\bibnamefont {Emplit}},\ and\ \bibinfo
  {author} {\bibfnamefont {M.}~\bibnamefont {Haelterman}},\ }\href
  {https://doi.org/10.1103/PhysRevLett.83.2328} {\bibfield  {journal} {\bibinfo
   {journal} {Physical Review Letters}\ }\textbf {\bibinfo {volume} {83}},\
  \bibinfo {pages} {2328} (\bibinfo {year} {1999})},\ \bibinfo {note}
  {publisher: American Physical Society}\BibitemShut {NoStop}%
\bibitem [{\citenamefont {Luo}\ \emph {et~al.}(2015)\citenamefont {Luo},
  \citenamefont {Xu}, \citenamefont {Erkintalo},\ and\ \citenamefont
  {Murdoch}}]{luo_resonant_2015}%
  \BibitemOpen
  \bibfield  {author} {\bibinfo {author} {\bibfnamefont {K.}~\bibnamefont
  {Luo}}, \bibinfo {author} {\bibfnamefont {Y.}~\bibnamefont {Xu}}, \bibinfo
  {author} {\bibfnamefont {M.}~\bibnamefont {Erkintalo}},\ and\ \bibinfo
  {author} {\bibfnamefont {S.~G.}\ \bibnamefont {Murdoch}},\ }\href
  {https://doi.org/10.1364/OL.40.000427} {\bibfield  {journal} {\bibinfo
  {journal} {Optics Letters}\ }\textbf {\bibinfo {volume} {40}},\ \bibinfo
  {pages} {427} (\bibinfo {year} {2015})}\BibitemShut {NoStop}%
\bibitem [{\citenamefont {Coen}\ and\ \citenamefont
  {Erkintalo}(2013)}]{coen_universal_2013}%
  \BibitemOpen
  \bibfield  {author} {\bibinfo {author} {\bibfnamefont {S.}~\bibnamefont
  {Coen}}\ and\ \bibinfo {author} {\bibfnamefont {M.}~\bibnamefont
  {Erkintalo}},\ }\href {https://doi.org/10.1364/OL.38.001790} {\bibfield
  {journal} {\bibinfo  {journal} {Optics Letters}\ }\textbf {\bibinfo {volume}
  {38}},\ \bibinfo {pages} {1790} (\bibinfo {year} {2013})}\BibitemShut
  {NoStop}%
\bibitem [{\citenamefont {Haelterman}\ \emph {et~al.}(1992)\citenamefont
  {Haelterman}, \citenamefont {Trillo},\ and\ \citenamefont
  {Wabnitz}}]{haelterman_dissipative_1992}%
  \BibitemOpen
  \bibfield  {author} {\bibinfo {author} {\bibfnamefont {M.}~\bibnamefont
  {Haelterman}}, \bibinfo {author} {\bibfnamefont {S.}~\bibnamefont {Trillo}},\
  and\ \bibinfo {author} {\bibfnamefont {S.}~\bibnamefont {Wabnitz}},\ }\href
  {https://doi.org/10.1016/0030-4018(92)90367-Z} {\bibfield  {journal}
  {\bibinfo  {journal} {Optics Communications}\ }\textbf {\bibinfo {volume}
  {91}},\ \bibinfo {pages} {401} (\bibinfo {year} {1992})}\BibitemShut
  {NoStop}%
\bibitem [{\citenamefont {Coen}\ \emph {et~al.}(2013)\citenamefont {Coen},
  \citenamefont {Randle}, \citenamefont {Sylvestre},\ and\ \citenamefont
  {Erkintalo}}]{coen_modeling_2013}%
  \BibitemOpen
  \bibfield  {author} {\bibinfo {author} {\bibfnamefont {S.}~\bibnamefont
  {Coen}}, \bibinfo {author} {\bibfnamefont {H.~G.}\ \bibnamefont {Randle}},
  \bibinfo {author} {\bibfnamefont {T.}~\bibnamefont {Sylvestre}},\ and\
  \bibinfo {author} {\bibfnamefont {M.}~\bibnamefont {Erkintalo}},\ }\href
  {https://doi.org/10.1364/OL.38.000037} {\bibfield  {journal} {\bibinfo
  {journal} {Optics Letters}\ }\textbf {\bibinfo {volume} {38}},\ \bibinfo
  {pages} {37} (\bibinfo {year} {2013})}\BibitemShut {NoStop}%
\bibitem [{\citenamefont {Hendry}\ \emph {et~al.}(2019)\citenamefont {Hendry},
  \citenamefont {Garbin}, \citenamefont {Murdoch}, \citenamefont {Coen},\ and\
  \citenamefont {Erkintalo}}]{hendry_impact_2019-1}%
  \BibitemOpen
  \bibfield  {author} {\bibinfo {author} {\bibfnamefont {I.}~\bibnamefont
  {Hendry}}, \bibinfo {author} {\bibfnamefont {B.}~\bibnamefont {Garbin}},
  \bibinfo {author} {\bibfnamefont {S.~G.}\ \bibnamefont {Murdoch}}, \bibinfo
  {author} {\bibfnamefont {S.}~\bibnamefont {Coen}},\ and\ \bibinfo {author}
  {\bibfnamefont {M.}~\bibnamefont {Erkintalo}},\ }\href
  {https://doi.org/10.1103/PhysRevA.100.023829} {\bibfield  {journal} {\bibinfo
   {journal} {Physical Review A}\ }\textbf {\bibinfo {volume} {100}},\ \bibinfo
  {pages} {023829} (\bibinfo {year} {2019})}\BibitemShut {NoStop}%
\bibitem [{\citenamefont {Anderson}\ \emph {et~al.}(2019)\citenamefont
  {Anderson}, \citenamefont {Bouchand}, \citenamefont {Liu}, \citenamefont
  {Weng}, \citenamefont {Obrzud}, \citenamefont {Herr},\ and\ \citenamefont
  {Kippenberg}}]{anderson_photonic_2019}%
  \BibitemOpen
  \bibfield  {author} {\bibinfo {author} {\bibfnamefont {M.~H.}\ \bibnamefont
  {Anderson}}, \bibinfo {author} {\bibfnamefont {R.}~\bibnamefont {Bouchand}},
  \bibinfo {author} {\bibfnamefont {J.}~\bibnamefont {Liu}}, \bibinfo {author}
  {\bibfnamefont {W.}~\bibnamefont {Weng}}, \bibinfo {author} {\bibfnamefont
  {E.}~\bibnamefont {Obrzud}}, \bibinfo {author} {\bibfnamefont
  {T.}~\bibnamefont {Herr}},\ and\ \bibinfo {author} {\bibfnamefont {T.~J.}\
  \bibnamefont {Kippenberg}},\ }\href {http://arxiv.org/abs/1909.00022}
  {\bibfield  {journal} {\bibinfo  {journal} {arXiv:1909.00022 [physics]}\ }
  (\bibinfo {year} {2019})},\ \bibinfo {note} {arXiv: 1909.00022}\BibitemShut
  {NoStop}%
\bibitem [{\citenamefont {Kobayashi}\ \emph {et~al.}(1988)\citenamefont
  {Kobayashi}, \citenamefont {Yao}, \citenamefont {Amano}, \citenamefont
  {Fukushima}, \citenamefont {Morimoto},\ and\ \citenamefont
  {Sueta}}]{kobayashi_optical_1988}%
  \BibitemOpen
  \bibfield  {author} {\bibinfo {author} {\bibfnamefont {T.}~\bibnamefont
  {Kobayashi}}, \bibinfo {author} {\bibfnamefont {H.}~\bibnamefont {Yao}},
  \bibinfo {author} {\bibfnamefont {K.}~\bibnamefont {Amano}}, \bibinfo
  {author} {\bibfnamefont {Y.}~\bibnamefont {Fukushima}}, \bibinfo {author}
  {\bibfnamefont {A.}~\bibnamefont {Morimoto}},\ and\ \bibinfo {author}
  {\bibfnamefont {T.}~\bibnamefont {Sueta}},\ }\href
  {https://doi.org/10.1109/3.135} {\bibfield  {journal} {\bibinfo  {journal}
  {IEEE Journal of Quantum Electronics}\ }\textbf {\bibinfo {volume} {24}},\
  \bibinfo {pages} {382} (\bibinfo {year} {1988})}\BibitemShut {NoStop}%
\bibitem [{\citenamefont {Lilienfein}\ \emph {et~al.}(2019)\citenamefont
  {Lilienfein}, \citenamefont {Hofer}, \citenamefont {Högner}, \citenamefont
  {Saule}, \citenamefont {Trubetskov}, \citenamefont {Pervak}, \citenamefont
  {Fill}, \citenamefont {Riek}, \citenamefont {Leitenstorfer}, \citenamefont
  {Limpert}, \citenamefont {Krausz},\ and\ \citenamefont
  {Pupeza}}]{lilienfein_temporal_2019}%
  \BibitemOpen
  \bibfield  {author} {\bibinfo {author} {\bibfnamefont {N.}~\bibnamefont
  {Lilienfein}}, \bibinfo {author} {\bibfnamefont {C.}~\bibnamefont {Hofer}},
  \bibinfo {author} {\bibfnamefont {M.}~\bibnamefont {Högner}}, \bibinfo
  {author} {\bibfnamefont {T.}~\bibnamefont {Saule}}, \bibinfo {author}
  {\bibfnamefont {M.}~\bibnamefont {Trubetskov}}, \bibinfo {author}
  {\bibfnamefont {V.}~\bibnamefont {Pervak}}, \bibinfo {author} {\bibfnamefont
  {E.}~\bibnamefont {Fill}}, \bibinfo {author} {\bibfnamefont {C.}~\bibnamefont
  {Riek}}, \bibinfo {author} {\bibfnamefont {A.}~\bibnamefont {Leitenstorfer}},
  \bibinfo {author} {\bibfnamefont {J.}~\bibnamefont {Limpert}}, \bibinfo
  {author} {\bibfnamefont {F.}~\bibnamefont {Krausz}},\ and\ \bibinfo {author}
  {\bibfnamefont {I.}~\bibnamefont {Pupeza}},\ }\href
  {https://doi.org/10.1038/s41566-018-0341-y} {\bibfield  {journal} {\bibinfo
  {journal} {Nature Photonics}\ ,\ \bibinfo {pages} {1}} (\bibinfo {year}
  {2019})}\BibitemShut {NoStop}%
\bibitem [{\citenamefont {Liu}\ \emph {et~al.}(2016)\citenamefont {Liu},
  \citenamefont {Brasch}, \citenamefont {Pfeiffer}, \citenamefont {Kordts},
  \citenamefont {Kamel}, \citenamefont {Guo}, \citenamefont {Geiselmann},\ and\
  \citenamefont {Kippenberg}}]{liu_frequency-comb-assisted_2016}%
  \BibitemOpen
  \bibfield  {author} {\bibinfo {author} {\bibfnamefont {J.}~\bibnamefont
  {Liu}}, \bibinfo {author} {\bibfnamefont {V.}~\bibnamefont {Brasch}},
  \bibinfo {author} {\bibfnamefont {M.~H.~P.}\ \bibnamefont {Pfeiffer}},
  \bibinfo {author} {\bibfnamefont {A.}~\bibnamefont {Kordts}}, \bibinfo
  {author} {\bibfnamefont {A.~N.}\ \bibnamefont {Kamel}}, \bibinfo {author}
  {\bibfnamefont {H.}~\bibnamefont {Guo}}, \bibinfo {author} {\bibfnamefont
  {M.}~\bibnamefont {Geiselmann}},\ and\ \bibinfo {author} {\bibfnamefont
  {T.~J.}\ \bibnamefont {Kippenberg}},\ }\href
  {https://doi.org/10.1364/OL.41.003134} {\bibfield  {journal} {\bibinfo
  {journal} {Optics Letters}\ }\textbf {\bibinfo {volume} {41}},\ \bibinfo
  {pages} {3134} (\bibinfo {year} {2016})}\BibitemShut {NoStop}%
\bibitem [{\citenamefont {Guo}\ \emph {et~al.}(2017)\citenamefont {Guo},
  \citenamefont {Karpov}, \citenamefont {Lucas}, \citenamefont {Kordts},
  \citenamefont {Pfeiffer}, \citenamefont {Brasch}, \citenamefont {Lihachev},
  \citenamefont {Lobanov}, \citenamefont {Gorodetsky},\ and\ \citenamefont
  {Kippenberg}}]{guo_universal_2017}%
  \BibitemOpen
  \bibfield  {author} {\bibinfo {author} {\bibfnamefont {H.}~\bibnamefont
  {Guo}}, \bibinfo {author} {\bibfnamefont {M.}~\bibnamefont {Karpov}},
  \bibinfo {author} {\bibfnamefont {E.}~\bibnamefont {Lucas}}, \bibinfo
  {author} {\bibfnamefont {A.}~\bibnamefont {Kordts}}, \bibinfo {author}
  {\bibfnamefont {M.~H.~P.}\ \bibnamefont {Pfeiffer}}, \bibinfo {author}
  {\bibfnamefont {V.}~\bibnamefont {Brasch}}, \bibinfo {author} {\bibfnamefont
  {G.}~\bibnamefont {Lihachev}}, \bibinfo {author} {\bibfnamefont {V.~E.}\
  \bibnamefont {Lobanov}}, \bibinfo {author} {\bibfnamefont {M.~L.}\
  \bibnamefont {Gorodetsky}},\ and\ \bibinfo {author} {\bibfnamefont {T.~J.}\
  \bibnamefont {Kippenberg}},\ }\href {https://doi.org/10.1038/nphys3893}
  {\bibfield  {journal} {\bibinfo  {journal} {Nature Physics}\ }\textbf
  {\bibinfo {volume} {13}},\ \bibinfo {pages} {94} (\bibinfo {year}
  {2017})}\BibitemShut {NoStop}%
\bibitem [{\citenamefont {Pfeiffer}\ \emph {et~al.}(2017)\citenamefont
  {Pfeiffer}, \citenamefont {Herkommer}, \citenamefont {Liu}, \citenamefont
  {Guo}, \citenamefont {Karpov}, \citenamefont {Lucas}, \citenamefont
  {Zervas},\ and\ \citenamefont {Kippenberg}}]{pfeiffer_octave-spanning_2017}%
  \BibitemOpen
  \bibfield  {author} {\bibinfo {author} {\bibfnamefont {M.~H.~P.}\
  \bibnamefont {Pfeiffer}}, \bibinfo {author} {\bibfnamefont {C.}~\bibnamefont
  {Herkommer}}, \bibinfo {author} {\bibfnamefont {J.}~\bibnamefont {Liu}},
  \bibinfo {author} {\bibfnamefont {H.}~\bibnamefont {Guo}}, \bibinfo {author}
  {\bibfnamefont {M.}~\bibnamefont {Karpov}}, \bibinfo {author} {\bibfnamefont
  {E.}~\bibnamefont {Lucas}}, \bibinfo {author} {\bibfnamefont
  {M.}~\bibnamefont {Zervas}},\ and\ \bibinfo {author} {\bibfnamefont {T.~J.}\
  \bibnamefont {Kippenberg}},\ }\href {https://doi.org/10.1364/OPTICA.4.000684}
  {\bibfield  {journal} {\bibinfo  {journal} {Optica}\ }\textbf {\bibinfo
  {volume} {4}},\ \bibinfo {pages} {684} (\bibinfo {year} {2017})}\BibitemShut
  {NoStop}%
\bibitem [{\citenamefont {Kelly}(1992)}]{kelly_characteristic_1992}%
  \BibitemOpen
  \bibfield  {author} {\bibinfo {author} {\bibfnamefont {S.~M.~J.}\
  \bibnamefont {Kelly}},\ }\href {https://doi.org/10.1049/el:19920508}
  {\bibfield  {journal} {\bibinfo  {journal} {Electronics Letters}\ }\textbf
  {\bibinfo {volume} {28}},\ \bibinfo {pages} {806} (\bibinfo {year}
  {1992})}\BibitemShut {NoStop}%
\bibitem [{\citenamefont {Hickstein}\ \emph {et~al.}(2018)\citenamefont
  {Hickstein}, \citenamefont {Kerber}, \citenamefont {Carlson}, \citenamefont
  {Chang}, \citenamefont {Westly}, \citenamefont {Srinivasan}, \citenamefont
  {Kowligy}, \citenamefont {Bowers}, \citenamefont {Diddams},\ and\
  \citenamefont {Papp}}]{hickstein_quasi-phase-matched_2018}%
  \BibitemOpen
  \bibfield  {author} {\bibinfo {author} {\bibfnamefont {D.~D.}\ \bibnamefont
  {Hickstein}}, \bibinfo {author} {\bibfnamefont {G.~C.}\ \bibnamefont
  {Kerber}}, \bibinfo {author} {\bibfnamefont {D.~R.}\ \bibnamefont {Carlson}},
  \bibinfo {author} {\bibfnamefont {L.}~\bibnamefont {Chang}}, \bibinfo
  {author} {\bibfnamefont {D.}~\bibnamefont {Westly}}, \bibinfo {author}
  {\bibfnamefont {K.}~\bibnamefont {Srinivasan}}, \bibinfo {author}
  {\bibfnamefont {A.}~\bibnamefont {Kowligy}}, \bibinfo {author} {\bibfnamefont
  {J.~E.}\ \bibnamefont {Bowers}}, \bibinfo {author} {\bibfnamefont {S.~A.}\
  \bibnamefont {Diddams}},\ and\ \bibinfo {author} {\bibfnamefont {S.~B.}\
  \bibnamefont {Papp}},\ }\href
  {https://doi.org/10.1103/PhysRevLett.120.053903} {\bibfield  {journal}
  {\bibinfo  {journal} {Physical Review Letters}\ }\textbf {\bibinfo {volume}
  {120}},\ \bibinfo {pages} {053903} (\bibinfo {year} {2018})},\ \bibinfo
  {note} {publisher: American Physical Society}\BibitemShut {NoStop}%
\bibitem [{\citenamefont {Kordts}\ \emph {et~al.}(2016)\citenamefont {Kordts},
  \citenamefont {Pfeiffer}, \citenamefont {Guo}, \citenamefont {Brasch},\ and\
  \citenamefont {Kippenberg}}]{kordts_higher_2016}%
  \BibitemOpen
  \bibfield  {author} {\bibinfo {author} {\bibfnamefont {A.}~\bibnamefont
  {Kordts}}, \bibinfo {author} {\bibfnamefont {M.~H.~P.}\ \bibnamefont
  {Pfeiffer}}, \bibinfo {author} {\bibfnamefont {H.}~\bibnamefont {Guo}},
  \bibinfo {author} {\bibfnamefont {V.}~\bibnamefont {Brasch}},\ and\ \bibinfo
  {author} {\bibfnamefont {T.~J.}\ \bibnamefont {Kippenberg}},\ }\href
  {https://doi.org/10.1364/OL.41.000452} {\bibfield  {journal} {\bibinfo
  {journal} {Optics Letters}\ }\textbf {\bibinfo {volume} {41}},\ \bibinfo
  {pages} {452} (\bibinfo {year} {2016})},\ \bibinfo {note} {publisher: Optical
  Society of America}\BibitemShut {NoStop}%
\bibitem [{\citenamefont {Huang}\ \emph {et~al.}(2017)\citenamefont {Huang},
  \citenamefont {Vinod}, \citenamefont {Yang}, \citenamefont {Yu},
  \citenamefont {Kwong},\ and\ \citenamefont
  {Wong}}]{huang_quasi-phase-matched_2017}%
  \BibitemOpen
  \bibfield  {author} {\bibinfo {author} {\bibfnamefont {S.-W.}\ \bibnamefont
  {Huang}}, \bibinfo {author} {\bibfnamefont {A.~K.}\ \bibnamefont {Vinod}},
  \bibinfo {author} {\bibfnamefont {J.}~\bibnamefont {Yang}}, \bibinfo {author}
  {\bibfnamefont {M.}~\bibnamefont {Yu}}, \bibinfo {author} {\bibfnamefont
  {D.-L.}\ \bibnamefont {Kwong}},\ and\ \bibinfo {author} {\bibfnamefont
  {C.~W.}\ \bibnamefont {Wong}},\ }\href {https://doi.org/10.1364/OL.42.002110}
  {\bibfield  {journal} {\bibinfo  {journal} {Optics Letters}\ }\textbf
  {\bibinfo {volume} {42}},\ \bibinfo {pages} {2110} (\bibinfo {year}
  {2017})}\BibitemShut {NoStop}%
\bibitem [{\citenamefont {Nielsen}\ \emph {et~al.}(2018)\citenamefont
  {Nielsen}, \citenamefont {Garbin}, \citenamefont {Coen}, \citenamefont
  {Murdoch},\ and\ \citenamefont {Erkintalo}}]{nielsen_invited_2018}%
  \BibitemOpen
  \bibfield  {author} {\bibinfo {author} {\bibfnamefont {A.~U.}\ \bibnamefont
  {Nielsen}}, \bibinfo {author} {\bibfnamefont {B.}~\bibnamefont {Garbin}},
  \bibinfo {author} {\bibfnamefont {S.}~\bibnamefont {Coen}}, \bibinfo {author}
  {\bibfnamefont {S.~G.}\ \bibnamefont {Murdoch}},\ and\ \bibinfo {author}
  {\bibfnamefont {M.}~\bibnamefont {Erkintalo}},\ }\href
  {https://doi.org/10.1063/1.5060123} {\bibfield  {journal} {\bibinfo
  {journal} {APL Photonics}\ }\textbf {\bibinfo {volume} {3}},\ \bibinfo
  {pages} {120804} (\bibinfo {year} {2018})}\BibitemShut {NoStop}%
\bibitem [{\citenamefont {Lucas}\ \emph {et~al.}(2017)\citenamefont {Lucas},
  \citenamefont {Guo}, \citenamefont {Jost}, \citenamefont {Karpov},\ and\
  \citenamefont {Kippenberg}}]{lucas_detuning-dependent_2017}%
  \BibitemOpen
  \bibfield  {author} {\bibinfo {author} {\bibfnamefont {E.}~\bibnamefont
  {Lucas}}, \bibinfo {author} {\bibfnamefont {H.}~\bibnamefont {Guo}}, \bibinfo
  {author} {\bibfnamefont {J.~D.}\ \bibnamefont {Jost}}, \bibinfo {author}
  {\bibfnamefont {M.}~\bibnamefont {Karpov}},\ and\ \bibinfo {author}
  {\bibfnamefont {T.~J.}\ \bibnamefont {Kippenberg}},\ }\href
  {https://doi.org/10.1103/PhysRevA.95.043822} {\bibfield  {journal} {\bibinfo
  {journal} {Physical Review A}\ }\textbf {\bibinfo {volume} {95}},\ \bibinfo
  {pages} {043822} (\bibinfo {year} {2017})}\BibitemShut {NoStop}%
\bibitem [{\citenamefont {Kim}\ \emph {et~al.}(2019)\citenamefont {Kim},
  \citenamefont {Okawachi}, \citenamefont {Jang}, \citenamefont {Yu},
  \citenamefont {Ji}, \citenamefont {Zhao}, \citenamefont {Joshi},
  \citenamefont {Lipson},\ and\ \citenamefont {Gaeta}}]{kim_turn-key_2019}%
  \BibitemOpen
  \bibfield  {author} {\bibinfo {author} {\bibfnamefont {B.~Y.}\ \bibnamefont
  {Kim}}, \bibinfo {author} {\bibfnamefont {Y.}~\bibnamefont {Okawachi}},
  \bibinfo {author} {\bibfnamefont {J.~K.}\ \bibnamefont {Jang}}, \bibinfo
  {author} {\bibfnamefont {M.}~\bibnamefont {Yu}}, \bibinfo {author}
  {\bibfnamefont {X.}~\bibnamefont {Ji}}, \bibinfo {author} {\bibfnamefont
  {Y.}~\bibnamefont {Zhao}}, \bibinfo {author} {\bibfnamefont {C.}~\bibnamefont
  {Joshi}}, \bibinfo {author} {\bibfnamefont {M.}~\bibnamefont {Lipson}},\ and\
  \bibinfo {author} {\bibfnamefont {A.~L.}\ \bibnamefont {Gaeta}},\ }\href
  {https://doi.org/10.1364/OL.44.004475} {\bibfield  {journal} {\bibinfo
  {journal} {Optics Letters}\ }\textbf {\bibinfo {volume} {44}},\ \bibinfo
  {pages} {4475} (\bibinfo {year} {2019})}\BibitemShut {NoStop}%
\bibitem [{\citenamefont {Li}\ \emph {et~al.}(2020)\citenamefont {Li},
  \citenamefont {Coen}, \citenamefont {Murdoch},\ and\ \citenamefont
  {Erkintalo}}]{li_experimental_2020}%
  \BibitemOpen
  \bibfield  {author} {\bibinfo {author} {\bibfnamefont {Z.}~\bibnamefont
  {Li}}, \bibinfo {author} {\bibfnamefont {S.}~\bibnamefont {Coen}}, \bibinfo
  {author} {\bibfnamefont {S.~G.}\ \bibnamefont {Murdoch}},\ and\ \bibinfo
  {author} {\bibfnamefont {M.}~\bibnamefont {Erkintalo}},\ }\href
  {http://arxiv.org/abs/2005.02995} {\bibfield  {journal} {\bibinfo  {journal}
  {arXiv:2005.02995 [physics]}\ } (\bibinfo {year} {2020})},\ \bibinfo {note}
  {arXiv: 2005.02995}\BibitemShut {NoStop}%
\bibitem [{\citenamefont {Pfeiffer}\ \emph {et~al.}(2018)\citenamefont
  {Pfeiffer}, \citenamefont {Herkommer}, \citenamefont {Liu}, \citenamefont
  {Morais}, \citenamefont {Zervas}, \citenamefont {Geiselmann},\ and\
  \citenamefont {Kippenberg}}]{pfeiffer_photonic_2018}%
  \BibitemOpen
  \bibfield  {author} {\bibinfo {author} {\bibfnamefont {M.~H.~P.}\
  \bibnamefont {Pfeiffer}}, \bibinfo {author} {\bibfnamefont {C.}~\bibnamefont
  {Herkommer}}, \bibinfo {author} {\bibfnamefont {J.}~\bibnamefont {Liu}},
  \bibinfo {author} {\bibfnamefont {T.}~\bibnamefont {Morais}}, \bibinfo
  {author} {\bibfnamefont {M.}~\bibnamefont {Zervas}}, \bibinfo {author}
  {\bibfnamefont {M.}~\bibnamefont {Geiselmann}},\ and\ \bibinfo {author}
  {\bibfnamefont {T.~J.}\ \bibnamefont {Kippenberg}},\ }\href
  {https://doi.org/10.1109/JSTQE.2018.2808258} {\bibfield  {journal} {\bibinfo
  {journal} {IEEE Journal of Selected Topics in Quantum Electronics}\ }\textbf
  {\bibinfo {volume} {24}},\ \bibinfo {pages} {1} (\bibinfo {year}
  {2018})}\BibitemShut {NoStop}%
\end{thebibliography}%

\end{document}